\documentclass[manuscript]{aastex}
\usepackage{hyperref}
\usepackage{graphicx}
\usepackage{bm}
\usepackage{booktabs}
\usepackage{color}

\newcommand{\Msun}{M_{\odot}}

\shortauthors{Pech et al.}
	
\begin{document}
\title{The Gould's Belt Very Large Array Survey V: The Perseus Region.}

\author{Gerardo Pech\altaffilmark{1}, Laurent Loinard\altaffilmark{1,2}, Sergio A. Dzib\altaffilmark{2}, Amy J. Mioduszewski\altaffilmark{3}, Luis F. Rodr\'iguez\altaffilmark{1,4},  Gisela N. Ortiz-Le\'on\altaffilmark{1}, Juana L. Rivera\altaffilmark{1}, Rosa M. Torres\altaffilmark{5}, Andrew F. Boden\altaffilmark{6}, Lee Hartman\altaffilmark{7}, Marina A. Kounkel\altaffilmark{7}, Neal J. Evans II\altaffilmark{8}, Cesar Brice\~no\altaffilmark{9}, John Tobin\altaffilmark{10} \and Luis A. Zapata\altaffilmark{1}}

\email{g.pech@crya.unam.mx}

\altaffiltext{1}{Instituto de Radioastronom\'ia y Astrof\'isica, Universidad Nacional Aut\'onoma de M\'exico, Apartado Postal 3-72, 58089 Morelia, Michoac\'an, M\'exico.}
\altaffiltext{2}{Max Planck Institut f\"ur Radioastronomie, Auf dem H\"ugel 69, D-53121 Bonn, Germany}
\altaffiltext{3}{National Radio Astronomy Observatory, Domenici Science Operations Center, 1003 Lopezville Road, Socorro, NM 87801, USA}
\altaffiltext{4}{King Abdulaziz University, P.O. Box 80203, Jeddah 21589, Saudi Arabia}
\altaffiltext{5}{Centro Universitario de Tonal\'a, Universidad de Guadalajara, Avenida Nuevo Perif\'erico No. 555, Ejido San Jos\'e Tatepozco, C.P. 48525, Tonal\'a, Jalisco, Mexico.}
\altaffiltext{6}{Division of Physics, Math and Astronomy, California Institute of Technology, 1200 East California Boulevard, Pasadena, CA 91125, USA}
\altaffiltext{7}{Department of Astronomy, University of Michigan, 500 Church Street, Ann Arbor, MI 48105, USA}
\altaffiltext{8}{Department of Astronomy, The University of Texas at Austin, 1 University Station, C1400, Austin, TX 78712, USA}
\altaffiltext{9}{Cerro Tololo Interamerican Observatory, Casilla 603, La Serena, Chile}
\altaffiltext{10}{Leiden Observatory, Leiden University, P.O. Box 9513, 2300 RA Leiden, The Netherlands}

\clearpage
\begin{abstract}
We present multi-epoch, large-scale ($\boldsymbol{\sim}$ 2000 arcmin${}^2$), fairly deep ($\sim$ 16 $\mu$Jy), high-resolution ($\sim$ 1 $\arcsec$) radio observations of  the Perseus star-forming complex obtained with the Karl G. Jansky Very Large Array at frequencies of 4.5 GHz and 7.5 GHz. These observations were mainly focused on the clouds NGC 1333 and IC 348, although we also observed several fields in other parts of the Perseus complex.  We detect a total of 206 sources, 42 of which are associated with young stellar objects (YSOs).  The radio properties of about 60\% of the YSOs are compatible with a non-thermal radio emission origin.  Based on our sample, we find a fairly clear relation between the prevalence of non-thermal radio emission and evolutionary status of the YSOs. By comparing our results with previously reported X-ray observations, we show that YSOs in Perseus follow a G\"udel-Benz relation with $\kappa$ = 0.03 consistent with other regions of star formation.  We argue that most of the sources detected in our observations but not associated with known YSOs are extragalactic, but provide a list of 20 unidentified radio sources whose radio properties are consistent with being YSO candidates.
Finally we also detect 5 sources with extended emission features which can clearly be associated with radio galaxies.
\end{abstract}


\section{Introduction}

The Perseus molecular complex is part of the ring-like structure of molecular clouds known as Gould's Belt.  It is of particular interest because it contains several of the regions within about 300 pc of the sun that are most actively forming low and intermediate-mass stars \citep{2008hsf1.book..308B}.
In this paper we describe the results of deep, large-scale radio observations of the Perseus complex, particularly of NGC 1333 and IC 348.   NGC 1333 is currently the most active region of star formation in the Perseus molecular cloud (see \citealt{2008hsf1.book..346W} for a recent review). It is one of the best studied extremely young clusters of low to intermediate mass stars, and one of the most active sites of ongoing star formation within 500 pc of the sun.  The molecular mass contained in the NGC 1333 region is approximately 450~$\Msun$ \citep{1996A&A...306..935W}. The cluster contains about 150 young stars with a median age of about $10^6$ years and a total mass in stars of about 100~$\Msun$ \citep{2008hsf1.book..346W}.  Distance determinations locate NGC 1333 at aproximately 235 pc \citep{2008PASJ...60...37H}.
The IC 348 region, on the other hand, is particularly important because it is well surveyed at a variety of wavelengths and intermediate in nature between dense clusters and loose associations (see \citealt{2008hsf1.book..372H} for a recent review).  It contains at least 360 stellar members, with a median age of $\sim$2--3 Myr, but this number may be as high as $\sim$400 \citep{2007AJ....134..411M}  and a total mass in stars of about 80~$\Msun$. It contains at least 26 brown dwarfs as well as some protostars, Herbig-Haro objects and starless sub-millimeter cores.  Distance measurements to IC 348 agree on a value of 300 pc \citep{2008hsf1.book..372H}.  This would place NGC 1333 and IC 348 at somewhat different distances.

Different mechanisms are required to explain the radio emission of young stars. These are either thermal (bremsstrahlung from H II regions or shocks/jets, winds, accretion flows, etc.) or non-thermal (gyrosynchrotron radiation from flares).   Embedded Class I protostars are known to drive collimated thermal winds or jets, so they are usually detected as thermal bremsstrahlung sources.  For more massive stars, the radio emission can also originate from optically thick or thin compact HII regions (\citealt{1988ApJ...333..788H}; \citealt{1991ApJ...371..626E}; \citealt{2000ApJ...531..861G}). More evolved YSOs (Class III sources) often exhibit non-thermal emission, but this type of emission has also been detected in some Class II and Class I sources (\citealt{2007A&A...469..985F}; \citealt{2010ApJ...718..610D}; \citealt{2013A&A...552A..51D}).  This emission is typically believed to arise from gyro-synchroton radiation from mildly relativistic electrons gyrating in the magnetosphere of young stellar objects. Gyro-synchrotron emission is characterized by a high brightness temperature, usually a high level of variability, and often a negative spectral index and some level of circular polarization (\citealt{1991ApJ...383..280H}; \citealt{1995MNRAS.272..469H}; \citealt{1996ApJ...459..193G}). 

In this paper we present observations obtained with the Karl G. Jansky Very Large Array (VLA) of the Perseus molecluar cloud and continue the work by \cite{2013ApJ...775...63D,2015ApJ...801...91D}, \cite{2014ApJ...790...49K} and \cite{2015ApJ...805....9O}  to discuss the population of radio sources in star forming regions contained within Gould's Belt and identify adequate target candidates for VLBI observations as part of the Gould's Belt Distance Survey \citep{2011RMxAC..40..205L}. This survey is aimed at measuring the distance to about 200 young stars distributed across five regions in Gould's Belt (Ophiuchus, Taurus, Perseus, Serpens and Orion).  

\section{Observations}

The observations were collected with the VLA of the National Radio Astronomy Observatory in B and BnA configurations.  Two frequency sub-bands, each 1 GHz wide, and centered at 4.5 and 7.5 GHz, respectively, were recorded simultaneously.  The observations were obtained in three sessions, on 2011 06/13 March, 14/25 April and 01/02/10/19/22 May, typically separated from one another by a month. This dual frequency, multi-epoch strategy was chosen to enable the characterization of the spectral index and variability of the detected sources, and to help with the identification of the emission mechanisms.  

Our observations cover mainly the NGC 1333 and IC 348 star forming regions (see Figure \ref{fig:map}).  We mapped the NGC~1333 area using a mosaic of 13 VLA pointings and the IC~348 area using a mosaic of 27 VLA pointings. Additionally, 7 pointings were selected to cover regions associated with other dust clouds. The distribution of the individual pointings in NGC~1333 and IC~348 follows a somewhat irregular pattern chosen to optimize the compromise between uniform sensitivity and the inclusion of the largest possible number of known young stars (see Figure \ref{fig:map}).  The FWHM of the primary beam (i.e., the field of view) of the VLA has a diameter of 10$'$ at 4.5 GHz and 6$'$ at 7.5 GHz. As a consequence, and taking into account the overlap of the beams, the mosaic of NGC~1333 covers an area of $\sim$~432~arcmin$^2$ at 4.5~GHz and $\sim$~235~arcmin$^2$ at 7.5~GHz.  The area covered by the IC~348 mosaic is $\sim$~800~arcmin$^2$ at 4.5~GHz and $\sim$~475~arcmin$^2$ at 7.5~GHz. 
All the observing sessions were organized as follows.  The standard flux calibrator 3C 147 was first observed for $\sim$10 minutes.  We subsequently spent one minute on the phase calibrator J0336+3218 followed by a series of three target pointings, spending three minutes on each. This phase calibrator/target sequence was repeated until all target fields were observed.  Thus, three minutes were spent on each target field for each epoch.

All data sets were edited and calibrated in a standard fashion using the Common Astronomy Software Applications package (CASA).  Once calibrated, the data at each frequency were imaged (Stokes parameter \textit{I}) using the CASA task \texttt{clean}. The NGC~1333 and IC~348 mosaics were constructed by setting the \emph{imagermode} parameter to ``mosaic" in the \texttt{clean} task. In order to take into account the non-coplanarity of the baselines far from the phase center, we set the \emph{gridmode} parameter to ``widefield" with \emph{wprojplanes}~$=$~64.  We also corrected the images for the primary beam attenuation.  The \emph{weighting} parameter in \texttt{clean} was set to ``briggs" with \emph{robust} subparameter set to 0.0. For wide fields of view (such as those considered here), bandwidth smearing can become an issue, causing the peak flux of a point source to be reduced while conserving its integrated flux density.  This worsens for sources at large distances from the phase center. To avoid this effect, all our observations were imaged using multi-frequency-scale (\texttt{mode=`mfs'} in \texttt{clean}).  This algorithm projects different frequencies onto different points of the uv-plane \citep{2011A&A...532A..71R}.

The noise levels in NGC~1333 and IC~348 were uniform across the mosaics and of order 20 to 30 $\mu$Jy at both frequencies in the individual epochs (see Table \ref{tab:resolution}).  For the individual fields, noise levels of 35 to 50 $\mu$Jy were reached at both frequencies in the individual epochs (see Table \ref{tab:resolution}). The improved noise levels in the mosaics results from the overlap between the individual fields.  To produce images with improved sensitivity, the three epochs were combined, resulting in noise levels of 15 to 18 $\mu$Jy uniformly across the mosaics, and $\sim$~28~$\mu$Jy for the individual fields (see again Table \ref{tab:resolution}.) The synthesized beam was order of 1\arcsec~and is given explicitely for each epoch, region, and frequency in Table \ref{tab:resolution}.

\section{Results}\label{sec:results}

To identify sources in our observations we use the concatenation of the three epochs, which provides the highest sensitivity.  Sources were first searched for using an automated source identification. The automated source identification was made using the \texttt{find sources} function in the \texttt{casaviewer}. However, automated identification resulted in some false positive detections and failed to detect some real sources (particularly near the outer edges of the images).  Therefore a visual inspection of the mosaics was performed in order to verify or discard the sources found by the automated identification, and to identify sources not detected by the automated identification. The criteria used to consider a detection as firm were (1) sources with a reported conterpart and a flux larger than three times the $\sigma$ noise of the area, and (2) sources with a flux larger than five times the $\sigma$ noise of the area and without reported counterparts. This procedure was adopted to minimize the possibility of reporting a large noise fluctuation as a real source. 

Our source count may in principle be affected by the so-called ``clean bias" (\citealt{1997ApJ...475..479W}; \citealt{1998AJ....115.1693C}). This is the effect that the \texttt{clean} algorithm can cause artificial changes to source fluxes and apparent image noise levels, i.e., it subtracts flux from real sources and redristibutes it on top of noise peaks or sidelobes.  This clean bias is larger in images with poor uv-plane coverage. In order to check if  this bias affected our results, we constructed histograms of pixel values in the source-free regions of the mosaic maps of NGC~1333 and IC~348. These histograms are shown in Figure \ref{fig:rmshisto}. We can see in these figures that the noise in our radio maps is well fitted by a gaussian (normal) distribution, with no evidence for superimposed excesses (wings or bumps). The fact that the noise follows a normal distribution suggests that any bias that may exist in our final images is negligible. Also, given that the noise on our radio maps follows a gaussian distribution, we can estimate the expected number of false peak detections in the maps.   The probability that the value of a standard normal random variable $X$ will exceed $x$ is given by the complement of the standard normal cumulative distribution function $Q(x) = 1 - \phi(x)$.  The cumulative distribution function, $\phi$, is given by

\begin{displaymath}
\phi(x) = \frac{1}{2} \left[ 1+ \mathrm{erf} \left( \frac{x}{\sqrt{2}} \right) \right]~\mathrm{,}
\end{displaymath}

\noindent where $\mathrm{erf}$ is the error function given by

\begin{displaymath}
\mathrm{erf} = \frac{1}{\sqrt{\pi}} \int_{-x}^{x} e^{-t^2} dt~\mathrm{.}
\end{displaymath}

With this, the probability that any independent pixel (the number of independent pixels is given by the ratio of the observed area to the area of the synthesized beam) will have a value greater than 5$\sigma$ is $Q(5) \sim 3\times10^ {-7}$. Given the number of independent pixels in our maps, we expect about 0.3 false detections in the NGC~1333 mosaic, about 1 false detections in the IC~348 mosaic, and about 1.5 false detections in the seven individual fields. Thus, we only expect about three false detections in our entire data set.  This is very small compared with the number of detected sources (see below), and will have a negligible statistical effect on our interpretation of the data. 

Following the procedure outlined above, a total of 206 sources were detected, 74 sources corresponding to NGC~1333, 91 sources to IC~348, and 41 sources corresponding to the seven individual fields (Tables \ref{tab:sourNGC1333}, \ref{tab:sourIC348} and \ref{tab:soursingles} respectively).  Only 125 of the 206 sources were detected at both frequencies, the remaining 81 were detected only in the 4.5 GHz images.  To reflect the fact that these sources were found as part of the Gould's Belt Very Large Array Survey, a source with coordinates hhmmss.ss-ddmmss.s will be named GBS-VLA Jhhmmss.ss-ddmmss.s.    
The flux of each source at 4.5 and 7.5 GHz are given in column 3 and 5 of Tables \ref{tab:sourNGC1333}, \ref{tab:sourIC348}, and \ref{tab:soursingles}.  Three sources of uncertainties on the fluxes are included: (1) the error that results from the statistical noise in the images, (2) a systematic uncertainty of 5\% resulting from possible errors in the absolute flux calibration and (3) uncertainties introduced by absolute pointing errors of the primary beam of the VLA antennas, as described by \citet{2014ApJ...788..162D}.  An estimate of the radio spectral index of each source (given in column 7 of Tables \ref{tab:sourNGC1333}, \ref{tab:sourIC348}, and \ref{tab:soursingles}) was obtained from the fluxes measured in each sub-band (at 4.5 and 7.5 GHz).  To calculate the errors on the spectral indices, the three sources of errors on the flux at each frequency were added in quadrature and the final error was obtained using standard error propagation theory.  We are aware that this procedure to obtain errors on the spectral indices is a little conservative given that the two frequencies were recorded simultaneusly making the ratio of the two bands independent of the absolute flux uncertainty but we prefer to maintain it to make sure we do not underestimate the errors. 

Once the sources were identified in the concatenated images we visually searched for them in the images of the individual epochs.  An estimate of the level of variability of the sources was measured by comparing the fluxes measured at the three epochs.   Specifically, we calculated, for each source and at each frequency, the difference between the highest and lowest measured fluxes and normalized by the maximum flux.  We did not search for variability on sources with extended emission because sensitivity and UV coverage effects can produce spurious variations. The resulting values, expressed in percent, are given in columns 4 and 6 of Tables \ref{tab:sourNGC1333}, \ref{tab:sourIC348}, and \ref{tab:soursingles}. We will consider a source as highly variable if its variability at any frequency is $\geq$~50\% at a $3\sigma$ level.

Given the positions of the radio sources in the region mapped, the next step is to try to determine if they are associated with a previously cataloged object and, in case they are, determine the nature of that object.  We searched the literature for previous radio detections, and for counterparts at X-ray, optical, near and mid-infrared and sub-millimetric wavelengths.  The search was done in the SIMBAD Astronomical Database, and accessed all the major catalogues.
We consider a radio source associated with a counterpart at another wavelength if the separation between the two was below the combined uncertainties of the two data sets.  This was about $1\farcs5$ for the optical and infrared catalog, but could be significantly larger for some of the radio catalogs. For instance, the NRAO/VLA Sky Survey \citep{1998AJ....115.1693C} has a relatively lower resolution ($\theta~=$~45\arcsec~FWHM) so the separation between a detection in our observations and an associated NVSS source could be as large as 15\arcsec. Additionally, this low resolution implies that one detection in the NVSS may correspond to blended emission of two or more of our detections (we will see examples of this shortly). Based on this search, we found a total of 112 sources with previously known counterparts at any frequency.  Whitin these known counterparts we found 42 YSOs, 20 stars, and 8 extragalactic sources. Additionally we found 42 objects with a known counterpart at some wavelength, but no information on the type of object.  These are 25 radio sources, 12 infrared sources, 2 submillimeter sources and 7 X-ray sources\footnote{One of these sources is both a radio an X-ray source; Three of these sources are both a radio and infrared sources.} (see tables \ref{tab:countNGC1333}, \ref{tab:countIC348} and \ref{tab:countsingles}).  We argue that all 25 radio sources and most (or even all) submillimeter sources, infrared sources and X-ray sources are background objects. The remaining 94 sources are (to our knowledge) new detections not previously reported in the literature.  

We searched for circular polarization (we imaged the Stokes parameter \textit{V}) towards the detected sources located in the inner quarter (in area) of the primary beam of the VLA.  Further out, beam squint can produce artificial polarization signals so polarization measurements become unreliable. Only four souces showed a significant level of circular polarization; they are listed in Table \ref{tab:pol}.

\section{Discussion}

\subsection{YSOs and their general radio properties}

From all detected radio sources, 42 are associated with objects previously classified as YSOs and are listed in Table \ref{tab:YSO}.  Of this total, 27 objects (i.e.\ roughly 60\%) show either high-variability, a negative spectral index or polarization, characteristics that are suggestive of non-thermal emission.  Furthermore, considering that a non-thermal source with flux~$\ga$~200~$\mu$Jy can be detected in a few hours of VLBA observations, 9 of these 27 objects have a sufficient flux density to permit VLBI parallax measurements.  The remaining 15 sources are not highly variable and have a spectral index that does not conclusively suggest non-thermal emission.  The evolutionary status is known for 34 of the 42 YSOs detected, while 21 have a known spectral type.
It has been found in other regions that, on average, more evolved stars show radio properties that resemble a non-thermal origin, i.e., they are more variable and have more negative spectral indices (\citealt{2013ApJ...775...63D,2015ApJ...801...91D}; \citealt{2014ApJ...790...49K}; \citealt{2015ApJ...805....9O}). In Figure \ref{fig:classvsalpha} we plot the radio spectral index as a function of evolutionary status.  On average, we can observe a tendency for more evolved YSOs to have a smaller (i.e., more negative) spectral index.  This tendency is more noticeable for objects between Class 0 to Class II.  The mean spectral index for Class III objects appears to divert from this tendency, but within the uncertainties, our result is in good agreement with previous ones. It indicates that the dominant emission process changes from somewhat optically thick free-free emission to either thin free-free or gyrosynchrotron emission as the YSOs evolve (see \citealt{2013ApJ...775...63D}). 

In Figure \ref{fig:classvsvar} we plot variability as a function of evolutionary status.   There is a clear tendency for more evolved YSOs to be, on average, significantly more variable than the younger ones. This agrees with results found for other star forming regions and is consistent with the result described above for the spectral index, as non-thermal emitters are often strongly variable. 
Finally, we can see in Figure \ref{fig:clasvsflux} that in our sample, on average, the flux density does not show a clear tendency with evolutionary status; this appears to differ from previous results (see Figure 4 of \citealt{2015ApJ...801...91D} and Figure 5 of  \citealt{2013ApJ...775...63D}) where the radio flux appears to be higher for more evolved sources.

In Figure \ref{fig:fluxvsst} we plot the flux density of the YSOs as a function of their spectral type (see Table \ref{tab:YSO} for references).  Our detected  YSOs have spectral type M, K and G. There is only one source of type F, only one of type A and no type B detection.  We find little systematic variation of the flux with spectral type, although some M and K stars in our sample seem to have a higher than average flux. This would be consistent with the results reported by \citet{2015ApJ...801...91D} for the Taurus region.

\subsection{Background sources}

In our observations we find only 8 objects that are explicitly classified as background extragalactic sources.  They are all located in the NGC~1333 region and have been previously reported in VLA radio observations by \citet{1999ApJS..125..427R} who classified them as extragalactic sources based on their negative spectral indices.
We also have a large number of radio sources detected for the first time, or associated with unidentified previously reported sources at radio, IR, sub-mm, or X-ray frequencies.  We argue that most of these first detections are extragalactic. \cite{1991AJ....102.1258F} showed that the number of expected background sources at 5~GHz can be described by 

\begin{displaymath}
\left( \frac{N}{\textrm{arcmin}^2} \right) = 0.42 \pm 0.05 \left( \frac{S}{30 \mu \textrm{Jy}} \right)^{-1.18 \pm 0.19}
\end{displaymath}

\noindent where $N$ is the number of sources per arcmin${}^2$ with flux density $>$S ($\mu$Jy).  We will use this relation to examine the statistics of the sources detected.  We will concentrate on the core region of each of our two mosaics (for which we have a continous coverage at uniform sensitivity) and on the observations at 4.5~GHz, which are more appropiate than those at 7.5~GHz for extragalactic objects, since those usually have negative spectral indices.

As we said before, we considered the minimum flux for a new detection to be 5$\sigma$. Assuming a uniform noise of 16~$\mu$Jy the minimum flux of the detected sources is about 80~$\mu$Jy, so in the approximately 432~arcmin${}^2$ covered by our NGC~1333 observations the number of expected background souces is 57$\pm$7. In the 800~arcmin${}^2$ covered by our IC~348 observations, the number of expected background sources is 105$\pm$12.  In NGC~1333, there are 8 sources classified as extragalactic and 8 unclassified sources previously reported at radio/X-ray wavelengths (see Table \ref{tab:countNGC1333}). In addition, there are 34 new (and unclassified) sources in this region.  This suggest that all unidentified sources are background objects, since we would then have 50 background objects, compared with the expected 57~$\pm$~7.
In IC~348, there are no identified extragalactic sources, 13 unclassified radio/X-ray sources, 2 sub-mm sources (see Table \ref{tab:countIC348}), and 40 new (and unclassified) radio sources.  This adds up to 55 sources anticipated to be background objects,  and remains significantly smaller than the expected 105~$\pm$~12 background objects in the area covered by our observations.  We note, however, that 18 of our radio sources are associated with objects formally classified as stars. Many of these sources come from infrared observations (\citealt{2003A&A...409..147P}; \citealt{2003ApJ...597..555M}) and have only been detected once.  Their classification as stars is therefore somewhat uncertain.  An example of this is GBS-VLA J034421.76+320918.3, as we shall see in Section \ref{sec:extended}.

For the seven individual fields a different approach is needed, because the sensitivity is not uniform across the field. We follow \cite{1998ASPC..132..303A} who showed that the number of expected background sources inside a field of diameter $\theta_F$ can be expressed as

\begin{displaymath}
N = 1.4\left\{ 1 - \exp{\left[ -0.0066 \left( \frac{\theta_F}{\textrm{arcmin}} \right)^2 \left( \frac{\nu}{5~\textrm{GHz}} \right)^2 \right]} \right\} \left(\frac{S_0}{\textrm{mJy}}\right)^{-0.75} \left( \frac{\nu}{5~\textrm{GHz}} \right)^{-2.52} 
\end{displaymath}

\noindent  where $S_0$ is the detectable flux density threshold.   For the observations at 4.5~GHz the field size is $\theta_F$ = 15$\arcmin$ and the previous expression can be written as

\begin{displaymath}
N = 1.28 S_0^{-0.75}
\end{displaymath}

\noindent In our observations $S_0$ is of order of 0.140~mJy so we expect to detect $N = 39 \pm 6$ background sources in the seven fields observed. We find 19 unclassified radio or infrared sources and 20 new (and unclassified) radio sources in our observations of these fields. This suggests that all radio sources in this fields are extragalactic.

Combining those three cases, we expect 201~$\pm$~14 background sources in our observations. In comparison, there are a total of 94 new and unclassified radio sources in our observations, 8 extragalactic sources, and 42 (presumably extragalactic) radio/X-ray/IR/sub-mm sources.  This adds up to 144 sources, suggesting a slight deficit, mostly corresponding to the IC~348 region, and which might be in part explained by the misclassification of some galaxies as stars (Sect. \ref{sec:extended}).

As discussed in \citet{2015ApJ...801...91D} extragalactic radio sources tend to show little variability and to have negative spectral indices.  We use these characteristics to search for possible YSOs candidates among the 156 radio objects detected here that are not classified as YSOs or as extragalactic. We find that 20 objects in that sample are highly variable.  These sources are listed in table \ref{tab:candidates} and could be considered as previously unidentified YSO candidates, although, given the inherent uncertainties on the radio properties of different classes of active galactic nuclei, they could also be extragalactic sources.

\subsection{The Radio--X-Ray relation} 

\cite{1993ApJ...405L..63G} and \cite{1994A&A...285..621B} showed that the radio and X-ray emissions of magnetically active stars are correlated by a relation of the form

\begin{displaymath}
\frac{L_X}{L_R} = \kappa \cdot 10^{15.5 \pm 1}~[\textrm{Hz}]~\textrm{.}
\end{displaymath}

From the 42 young stars in our sample 22 have known X-ray luminosities and will be used to study the $L_X/L_R$ relation for YSOs.  X-ray luminosities were searched in NASA's High Energy Astrophysics Science Archive Research Center (HEASARC) and have been corrected to adopt a distance of 235~pc to the entire Perseus molecular cloud \citep{2008PASJ...60...37H}. In Figure \ref{fig:benzguedel}, we plot the X-ray luminosities of the YSOs as a function of their radio luminosities at both frequencies observed in this work (magenta symbols). For comparison, we also plot the results obtained in Ophiuchus (green symbols; from \citealt{2013ApJ...775...63D}), Serpens-W40 (red symbols; from \citealt{2015ApJ...805....9O}), Orion (yellow symbols; from \citealt{2014ApJ...790...49K}), and Taurus--Auriga (blue symbols; from \citealt{2015ApJ...801...91D}). We find that for our sample $L_X/L_R \le 10^{15.5}$, in agreement with results in other regions and also with those obtained by \cite{2004ApJ...613..393G} and \cite{2010ApJ...719..691F}.  A relation $L_X/L_R \approx 10^{14\pm 1}$ provides a good match to the distribution of points in this plot and is consistent with previous results of the Gould's Belt Very Large Array Survey.  The correlation, however is not very strong ($r$ $\sim$ 0.55). This is equivalent, in terms of the G\"udel-Benz relation to $\kappa=0.03$ for YSOs.  

\subsection{Comments on individual sources}
 
\subsubsection{Extended sources}\label{sec:extended}

In our observations, we find 5 sources with extended emission features, one in NGC~1333 and 4 in IC~348.  

In Figure \ref{fig:GBS032920} we show radio maps of source GBS-VLA J032920.67+311549.5 in NGC~1333 at 4.5 GHz and 7.5 GHz, corresponding to the concatenation of the three epochs.  This source is $0\farcs33$ from the previously cataloged radio source VLA 32 which was classified as extragalactic (\citealt{2011ApJ...736...25F}; \citealt{1999ApJS..125..427R}).  Based on the angular separation from GBS-VLA J032920.67+311549.5, it is likely that VLA 32 is the counterpart of our detection. This source is also associated with NVSS~032920+311549 reported by \citet{1998AJ....115.1693C} (the angular separation is only 1\farcs47), which has a flux at 1.4~GHz of 8.7 mJy.  

 All four extended sources in IC  348 are located within a square area of about $5\arcmin \times 5\arcmin$ approximately centered at R.A. $03^h 44^m 23^s$, DEC. $+32\degr ~11\arcmin~13\arcsec$. In Figures \ref{fig:IC348C} and \ref{fig:IC348X} we show radio maps of the region containing all these sources, corresponding to the concatenation of the three epochs.   
The source GBS-VLA~J034411.69+321039.4 has no previously reported counterpart in the SIMBAD catalog, and no associated source was found in the NVSS catalog.
The source GBS-VLA J034421.76+320918.3 is 1$\farcs$58 from previously cataloged source Cl*IC~348 MM 42 detected by \citet{2003ApJ...597..555M} in infrared observations of IC~348. The source Cl*IC~348~MM~42 is classified as a star, yet no spectral type nor SED classification is given for this source (\citealt{2003A&A...409..147P}; \citealt{2003ApJ...597..555M}).  To our knowledge, there is no other identification for Cl*IC~348~MM~42.  Based on their angular separation, it is likely that Cl*IC~348~MM~42 is the infrared counterpart of our detection, but if GBS-VLA J034421.76+320918.3 is actually a radio galaxy (as we will discuss later) these sources might be completely unrelated and just in the same direction as observed from the earth.   Alternatively, Cl*IC~348~MM~42 may have been erroneously identified as a star, being instead an extragalactic source. This source is also 0\farcs34 from radio source NVSS~034421+320918, which is reported to have an integrated flux at 1.4~GHz of 46.2~mJy \citep{1998AJ....115.1693C}. 

The source GBS-VLA J034433.04+321241.3 is 0$\farcs$52 from previously cataloged source CXOPZ 110 detected by \citet{2001AJ....122..866P} in Chandra observations.  To our knowledge, no optical or infrared counterpart has been detected for CXOPZ 110.  Based on its angular separation from GBS-VLA J034433.04+321241.3 we can assume CXOPZ~110 to be the counterpart of our detection.  This source is also 14\farcs9 from radio source NVSS~034433+321255 \citep{1998AJ....115.1693C}.

The source GBS-VLA 034433.91+321307.5 has no previously reported counterpart in the SIMBAD catalog.  This source is 13\farcs7 from radio source NVSS~034433+321255, with a reported flux at 1.4~GHz of 223.0~mJy \citep{1998AJ....115.1693C}. Based on the angular separation to GBS-VLA J034433.04+321241.3 and GBS-VLA 034433.91+321307.5, and the angular resolution of the NVSS observations ($\theta~=$~45\arcsec), we can assume that  sources GBS-VLA J034433.04+321241.3 and GBS-VLA 034433.91+321307.5 might have been detected, yet not resolved, by \citet{1998AJ....115.1693C} and that the NVSS~034433+321255 emission would encompass the combined emission of both of our sources.
It is not completely clear from Figure \ref{fig:IC348C} if the emission feature observed directly southeast of GBS-VLA J034433.91+321307.5 can actually be associated to this source or if it might correspond to a northeast feature of the source GBS-VLA J034433.04+321241.3, further observations are necessary to clarify this.

As we can observe in the Figures, the sources mentioned above exhibit a double and fairly symmetrical structure acompanied by some fainter, more elongated and collimated jet like emission.   All features are clearly more luminous in the 4.5 GHz map than in the 7.5 GHz, and even brighter in the NVSS 1.4~GHz data.  This is consistent with all sources having negative spectral indices which suggests non-thermal emission, most probably synchrotron.  Relativistic beaming of the synchroton emission can explain why the lobes and jet like emission are observed stronger and more defined in one direction than the other.  Based on these characteristics we propose that these sources correspond to at least 4 radio galaxies.
These radio galaxies are at least $10\degr$ away from the known Perseus Cluster so we can assume they are not members of this cluster.  

\subsubsection{GBS-VLA J032903.75+311603.7}

In Figure \ref{fig:VLA4} we show a radio map of source GBS-VLA J032903.75+311603.7. It is $0\farcs47$ away from previously cataloged radio source VLA 4b and $0\farcs63$ away from previously cataloged radio source VLA 4a.  The sources VLA 4a and VLA 4b form a close binary separated by $0\farcs3$ or $\sim65$ AU (\citealt{2004ApJ...605L.137A}; \citealt{2000ApJ...542L.123A}). The source VLA 4b exhibits stronger mm emission than VLA 4a which suggests the source is associated with a larger amount of dust, probably a circumstellar dust disk, while the source VLA 4a appears to be the counterpart of the optically visible star SVS 13.  As we can see from the Figure, we were not able to resolve the VLA 4a/4b binarity, detecting a single source instead.

\subsection{Proper motions of YSOs in Perseus}

A previous and accurate analysis of proper motions for young stellar objects in NGC~1333 has been made by \citet{2008AJ....136.2238C}. They used Very Large Array data taken over 10 years, ranging from 1989 to 1999, to measure the proper motions of four sources in NGC~1333 (VLA 2, VLA 3, VLA 4a, VLA 4b, see their Table 2; also their Figure 2).  They found average values of $\mu_\alpha \cos(\delta) = 9 \pm 1$ mas~yr${}^{-1}$ and $\mu_\delta = -10 \pm 2$ mas~yr${}^{-1}$.   \citet{2011ApJ...736...25F}  presented Very Large Array radio observations of the NGC~1333 region obtained in 2006.  In their observations they also detected and reported positions of the sources studied by \citet{2008AJ....136.2238C} (see Table 6 and Table 7 of \citet{2011ApJ...736...25F}).  Sources VLA 2 and VLA 3 are counterparts for our sources GBS-VLA J032901.96+311538.1 and GBS-VLA J032903.38+311601.6.  We can use the position of these sources in our observations, along with the positions reported by \citet{2008AJ....136.2238C} and \citet{2011ApJ...736...25F} to expand the time span of the study of proper motions of these two sources.   Source GBS-VLA J032903.75+311603.7 corresponds to the binary source VLA 4a/b of \citet{2008AJ....136.2238C}, but as mentioned before we are not able to detect this binarity. This source also appears as a single source in the \citet{2011ApJ...736...25F} observations, so it is not possible to accurately include this source in our study.
Our results are shown in Figure \ref{fig:pmvla2} and Figure \ref{fig:pmvla3}.   We obtain averaged values of $\mu_\alpha \cos(\delta) = 6.74 \pm 0.67$ mas~yr${}^{-1}$ and $\mu_\delta = -15.32 \pm 0.92$ mas~yr${}^{-1}$ (see Table \ref{tab:motions}).   Our results are compatible with those reported by \citet{2008AJ....136.2238C} but are significantly more accurate. They might be taken as a better measure of the proper motion of the molecular cloud NGC~1333 as a whole. 

\section{CONCLUSIONS AND PERSPECTiVES}

We have presented a multi-epoch VLA survey at 4.5 GHz and 7.5 GHz of the NGC 1333 and IC 348 clouds along with seven individual fields in the Perseus star-forming region.  The multi-epoch, two frequency strategy has enabled us to determine the radio properties of the detected sources, and provided clues about the nature of their radio emission and the nature of the objects.  We detected a total of 206 sources (74 in NGC 1333, 91 in IC 348 and 41 in the seven individual fields), 42 of them related to YSOs.  Most of remaining sources are probably extragalactic  sources, but we provide a list of sources whose radio characteristics make them YSOs candidates. There is a clear tendency for the more evolved YSOs in our sample to exhibit radio properties consistent with a non-thermal origin.

Roughly 60\% of the YSOs have radio emission consistent with a non-thermal origin (gyrosynchrotron); at least 9 of these sources may constitute suitable targets for future VLBI observations.  By comparing our results with previous X-ray observations we found that the sources in Perseus follow the so-called G\"udel-Benz relation with $\kappa =$ 0.03, this is consistent with the results in other star-forming regions.

We also detected 5 sources with extended emission which can clearly be associated with radio galaxies. Several of these sources had been reported in the NVSS catalog of \citet{1998AJ....115.1693C}.

\begin{figure}
\begin{center}
\includegraphics{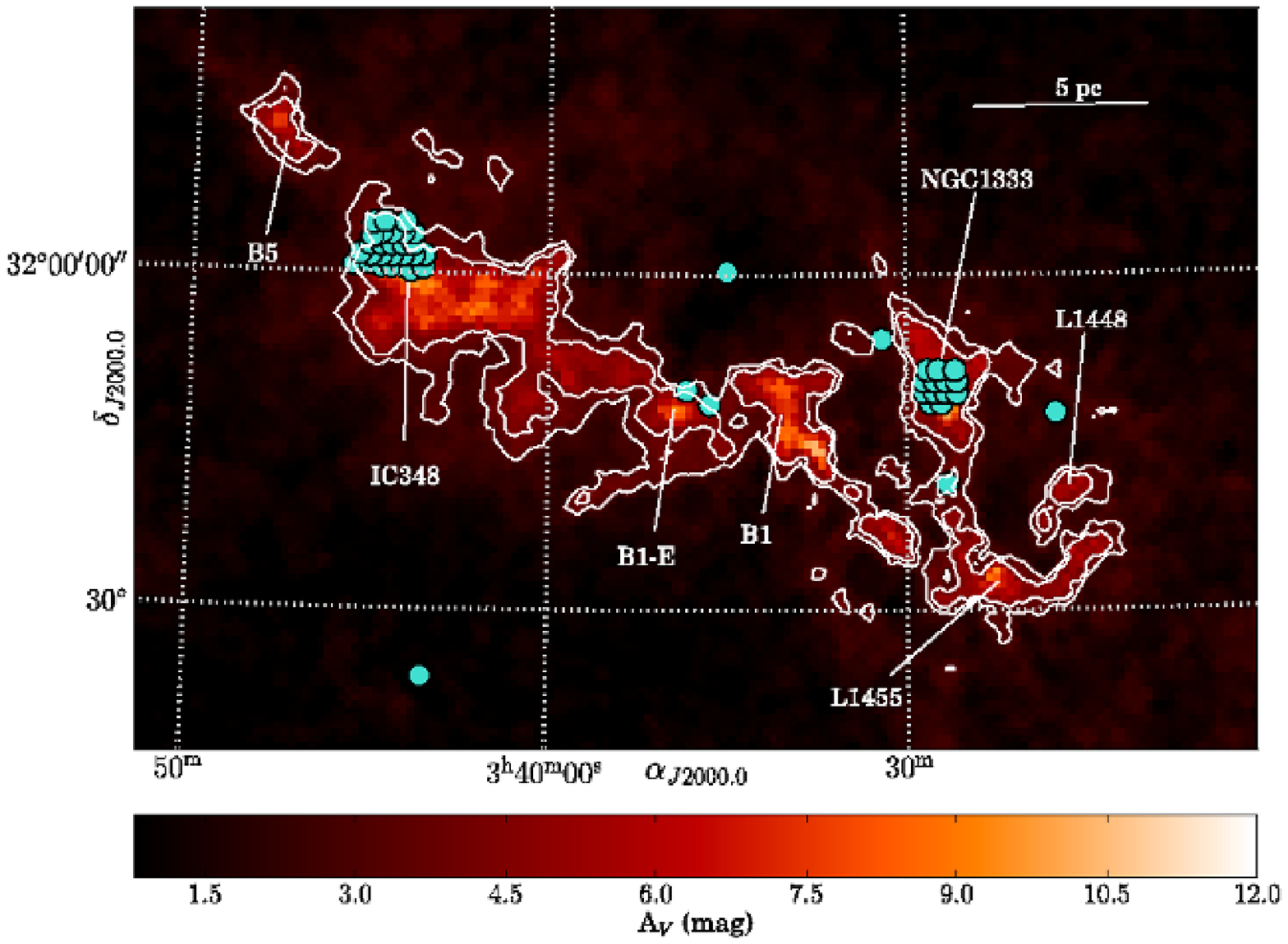}
\caption{Extinction map of the Perseus complex obtained as part of the COMPLETE project \citep{2006AJ....131.2921R} based on the 2MASS data \citep{2006AJ....131.1163S}. A linear distance is provided (assuming a distance of 235 pc to the entire region --- \citealt{2008PASJ...60...37H}). The turquoise circles indicate the areas mapped with the VLA for the survey presented here.  The diameter of each circle is $6\arcmin$ and corresponds to the primary beam of the VLA at 7.5~GHz. Note that the field of view, and therefore also the total mapped area, at 4.5~GHz is significantly larger. }
\label{fig:map}
\end{center}
\end{figure}

\begin{figure}
\begin{center}
\includegraphics{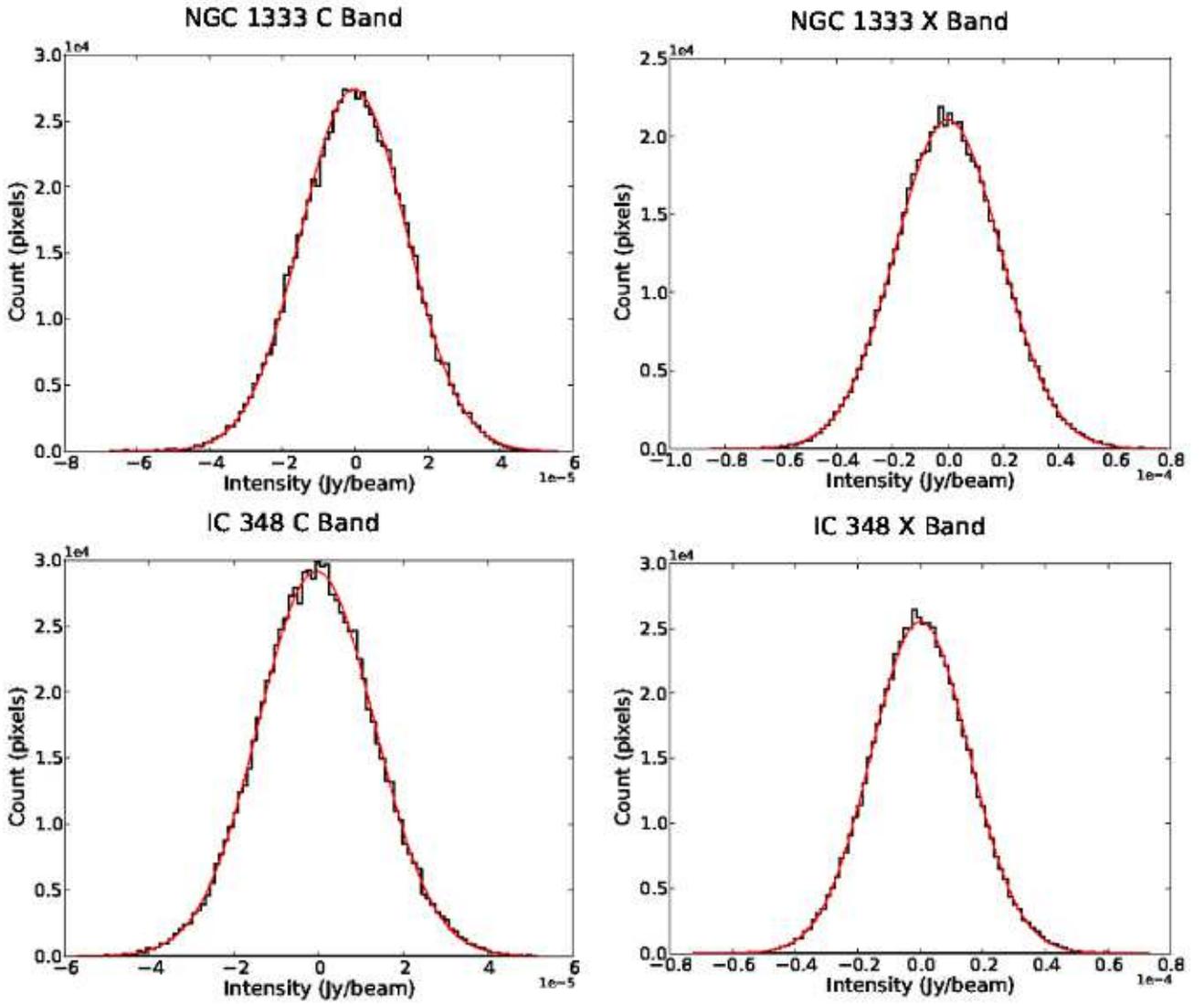}
\caption{Histograms of the pixel values of the noise in radio maps of NGC~1333 and IC~348 at 4.5~GHz and 7.5~GHz.  The red line is a gaussian fit to its respective histrogram.}
\label{fig:rmshisto}
\end{center}
\end{figure}

\begin{figure}[htbp]
\begin{center}
\includegraphics{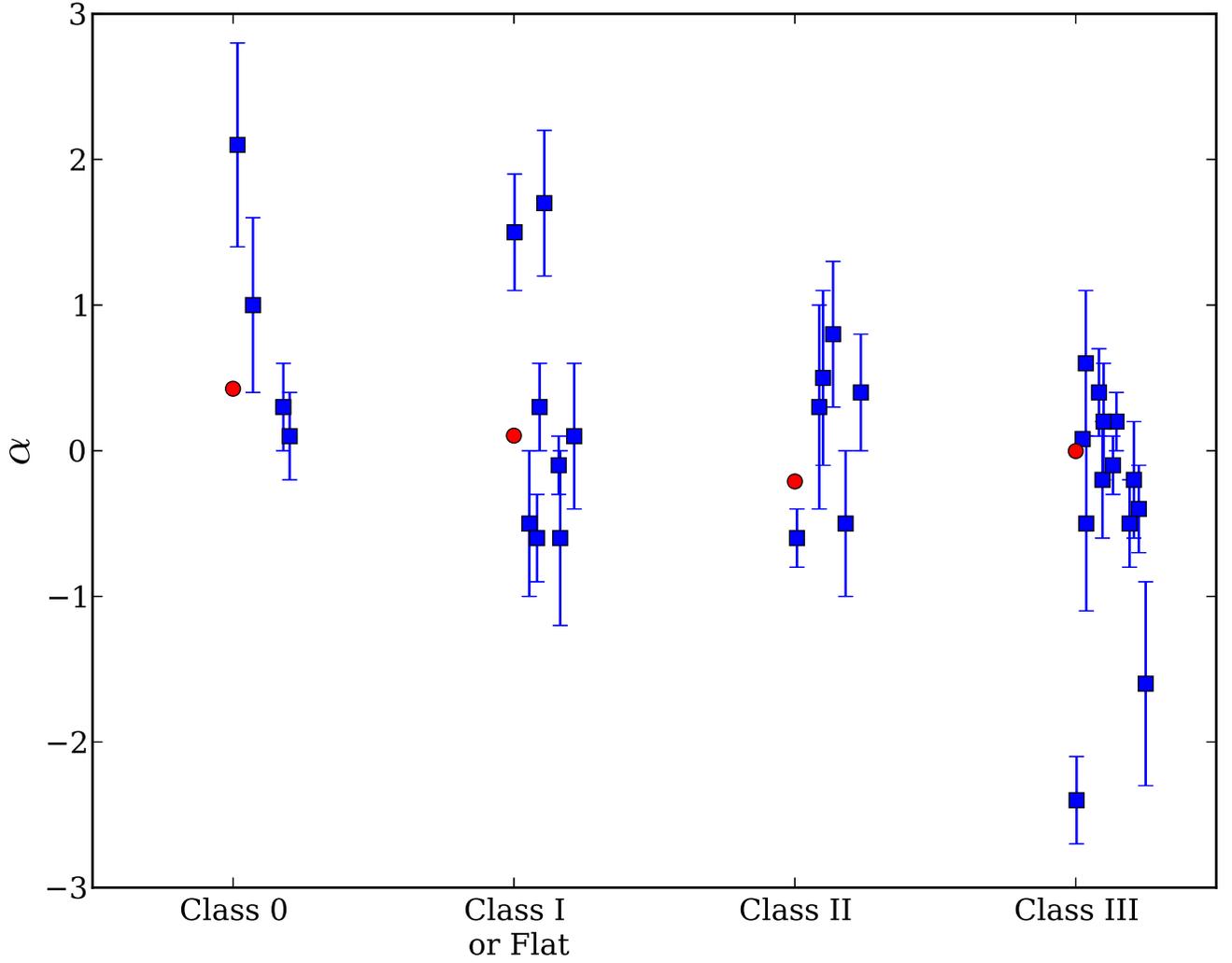}
\caption{Spectral index as a function of the YSO evolutionary stages for YSOs in Perseus.  The individual sources are shown with their error bars, and the red circles indicate the weighted average spectral index for each category.}
\label{fig:classvsalpha}
\end{center}
\end{figure}

\begin{figure}[htbp]
\begin{center}
\includegraphics{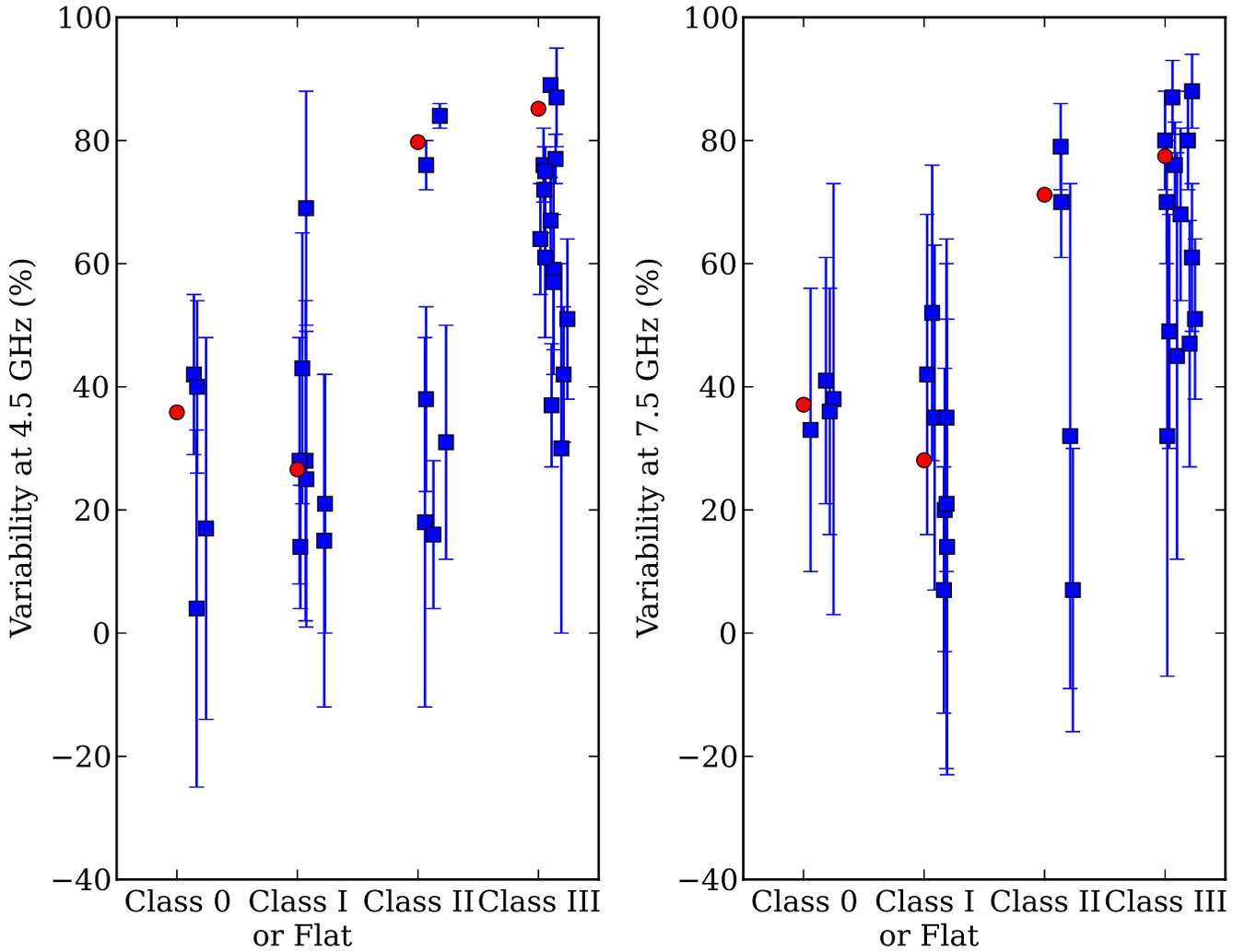}
\caption{Variability at 4.5 GHz (left) and 7.5 GHz (right) as a function of YSO evolutionary status. The individual sources are shown with their error bars, and the red circles indicate the weighted average variability for each category.}
\label{fig:classvsvar}
\end{center}
\end{figure}

\begin{figure}[htbp]
\begin{center}
\includegraphics{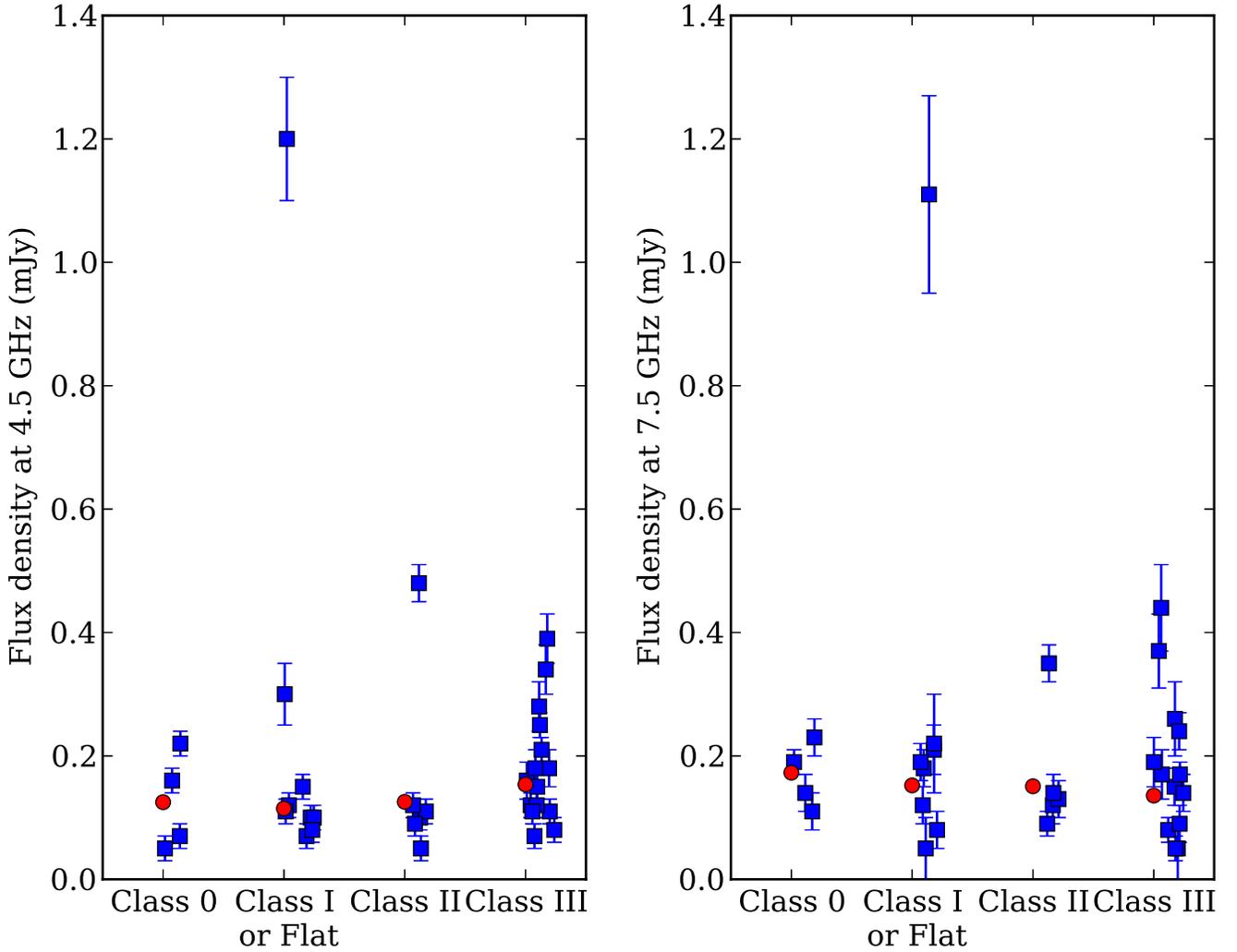}
\caption{Radio flux at 4.5 GHz (left) and 7.5 GHz (right) as a function of YSO evolutionary status. The individual sources are shown with their error bars, and the red circles indicate the weighted average flux for each category.}
\label{fig:clasvsflux}
\end{center}
\end{figure}

\begin{figure}[htbp]
\begin{center}
\includegraphics{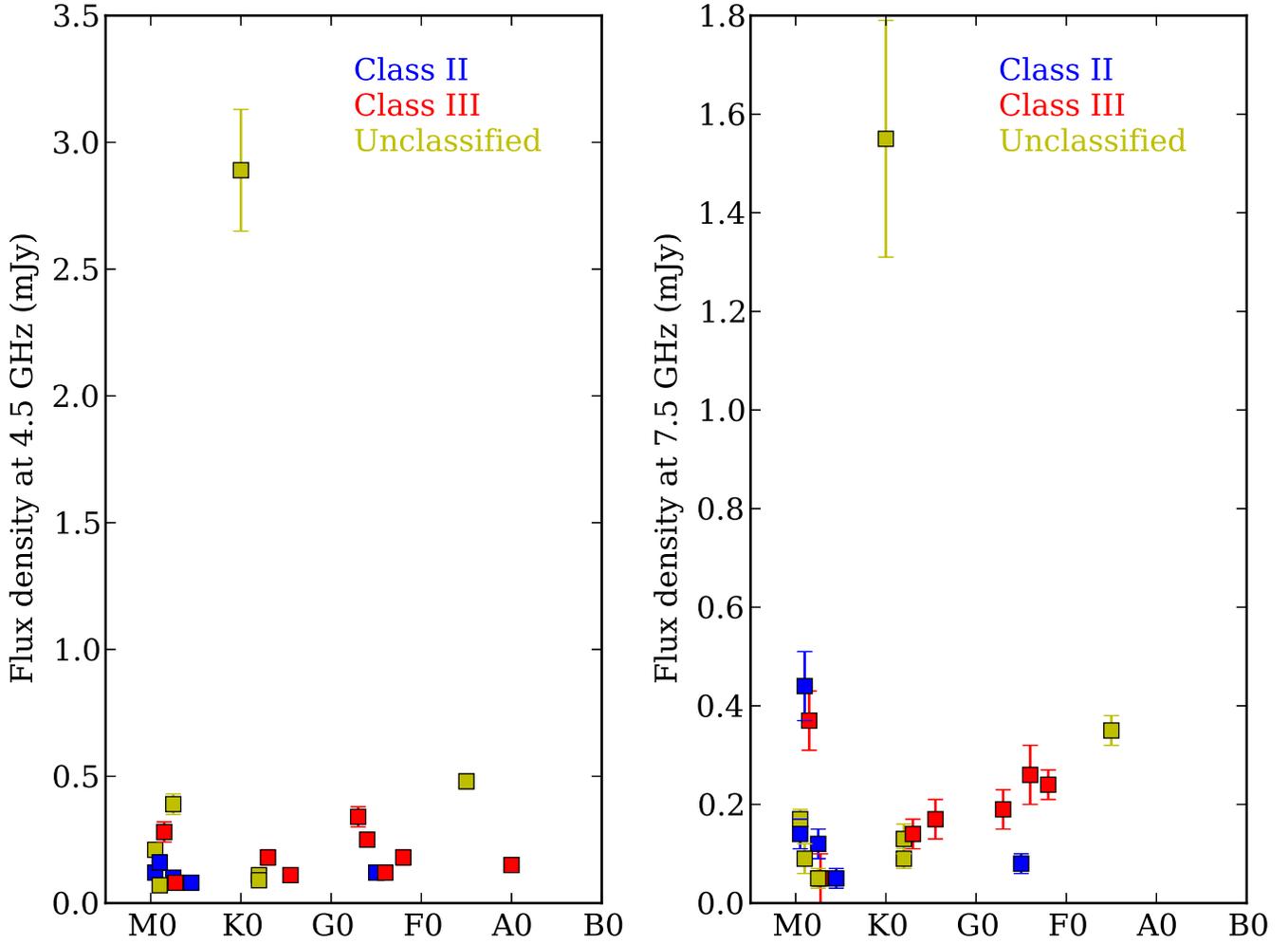}
\caption{Radio flux at 4.5 GHz (left) and 7.5 GHz (right) as a function of YSO spectral type. Colors indicate the evolutionary class of the object as listed at the top-right of the diagrams.}
\label{fig:fluxvsst}
\end{center}
\end{figure}

\begin{figure}[htbp]
\begin{center}
\includegraphics{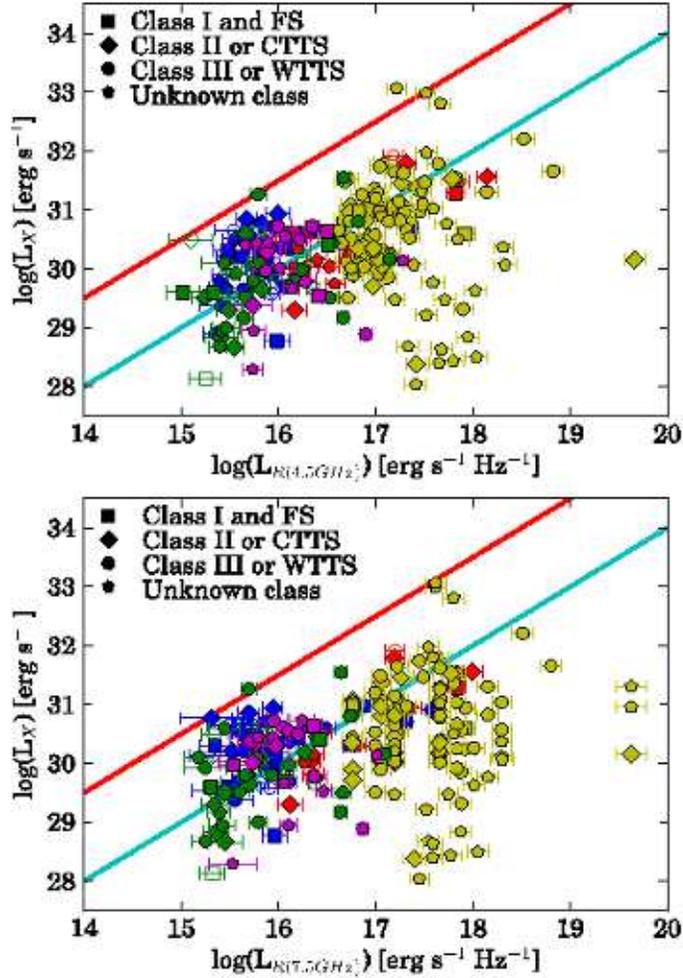}
\caption{X-ray luminosity as a function of radio luminosity.  The red line corresponds to the fiducial G\"udel-Benz relation with $\kappa=1$. The blue line corresponds to the G\"udel-Benz relation but with $\kappa=0.03$. Symbols indicate the evolutionary status of the object as explained at the top-left of the diagram.  Colors indicates YSOs in different star--forming regions: Perseus (magenta, this work), Ophiuchus (green symbols; from \citealt{2013ApJ...775...63D}), Serpens-W40 (red symbols; from \citealt{2015ApJ...805....9O}), Orion (yellow symbols; from \citealt{2014ApJ...790...49K}), and Taurus--Auriga (blue symbols; from \citealt{2015ApJ...801...91D}).  Open symbols indicate sources whose radio emission is thermal and solid symbols indicate non-thermal radio sources.}
\label{fig:benzguedel}
\end{center}
\end{figure}

\begin{figure}[htbp]
\begin{center}
\includegraphics{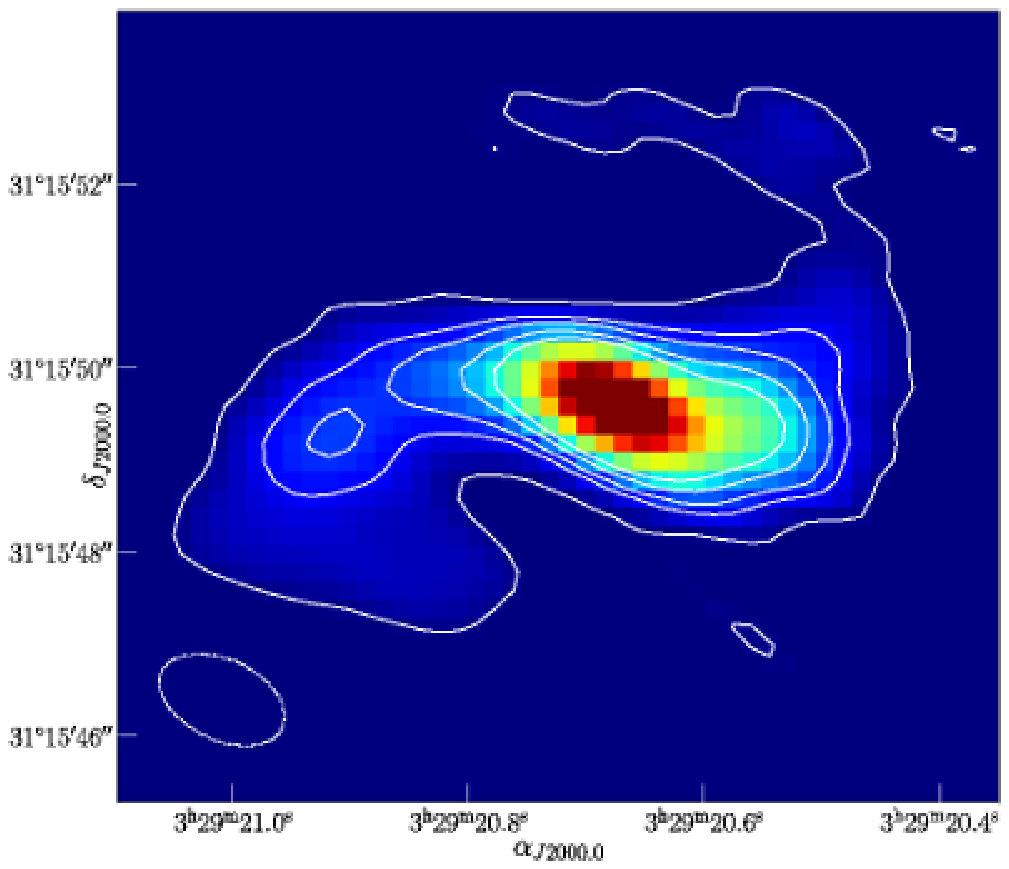}
\includegraphics{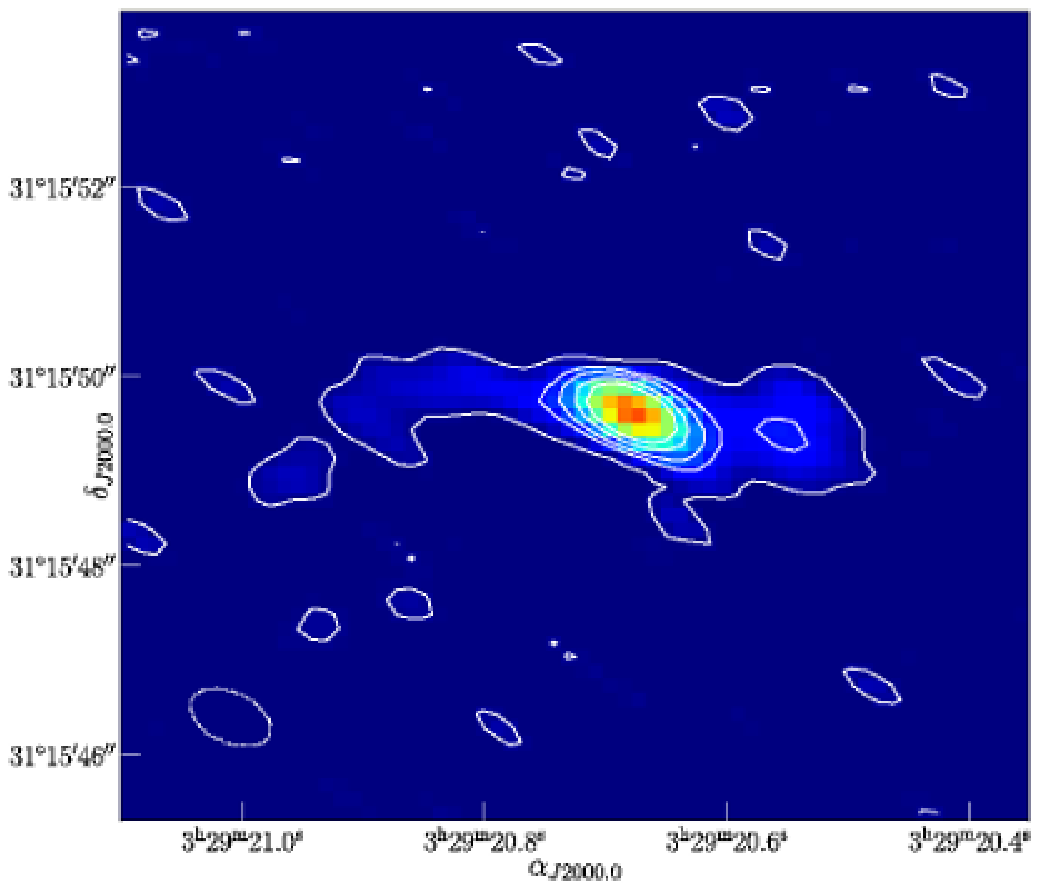}
\caption{(Top:) Radio map of source GBS--VLA J032920.67+311549.5 in NGC 1333 at 4.5 GHz. Contours are 2.5, 7, 10, 15, and 20 times the rms of the image (rms = 16~$\mu$Jy). The synthesized beam, shown in the bottom-left corner, is $1\farcs4 \times 0\farcs9$; P. A. = $65\degr$. 
(Bottom:) Radio map of source GBS--VLA J032920.67+311549.5 at 7.5 GHz. Contours are 2.5, 7, 10, 15 and 20 times the rms of the image (rms = 18~$\mu$Jy). The synthesized beam, shown in the bottom-left corner, is $0\farcs9 \times 0\farcs5$; P. A. = $66\degr$.}
\label{fig:GBS032920}
\end{center}
\end{figure}

\begin{figure}[htbp]
\begin{center}
\includegraphics{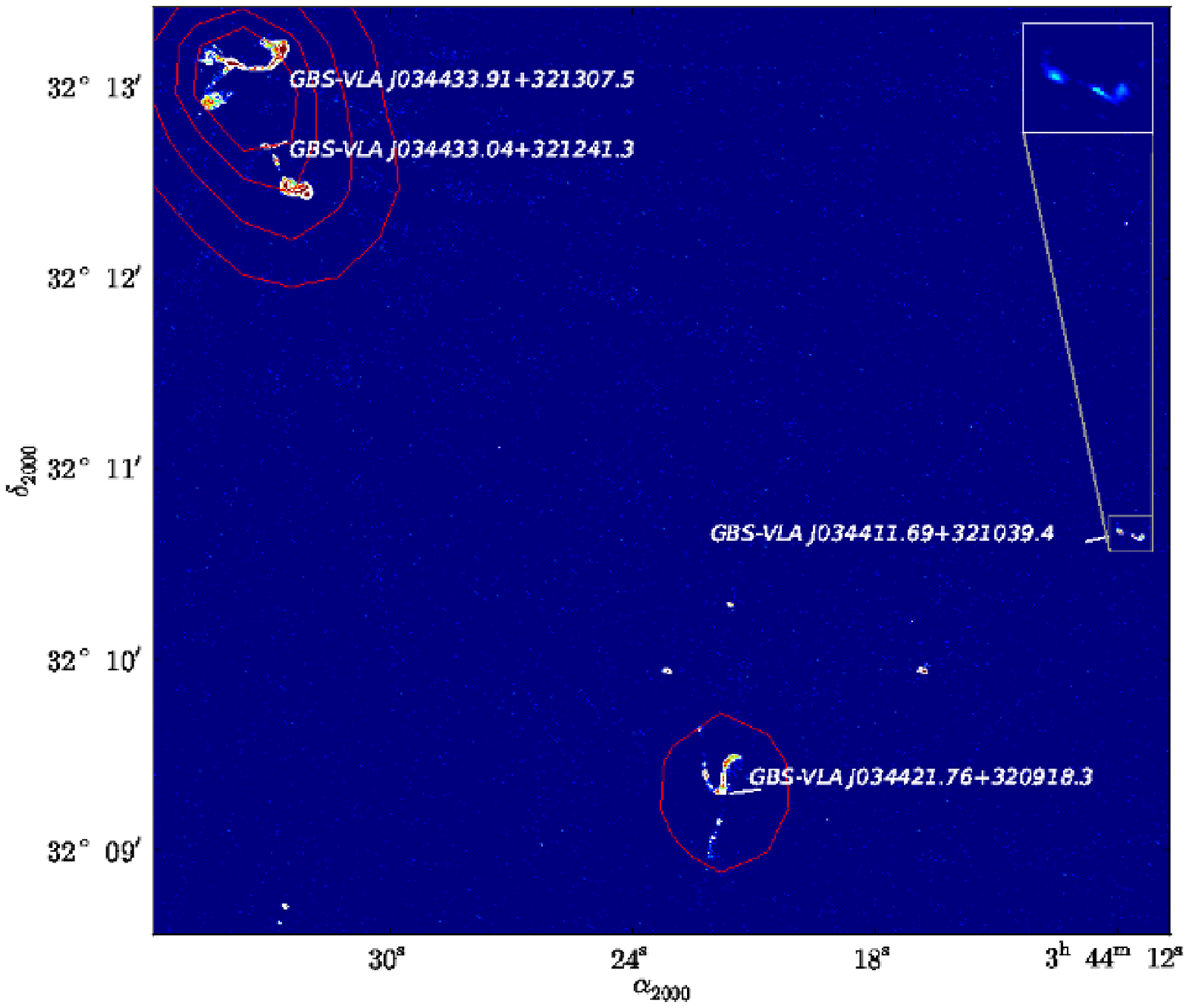}
\caption{Radio map showing the extended sources found in IC 348 observations at 4.5 GHz overlaid with 1.4~GHz NVSS observations (red contours; \citealt{1998AJ....115.1693C}). The white contours are 4, 10, 20 and 35 times the rms of the image (rms = 16~$\mu$Jy).  }
\label{fig:IC348C}
\end{center}
\end{figure}

\begin{figure}[htbp]
\begin{center}
\includegraphics{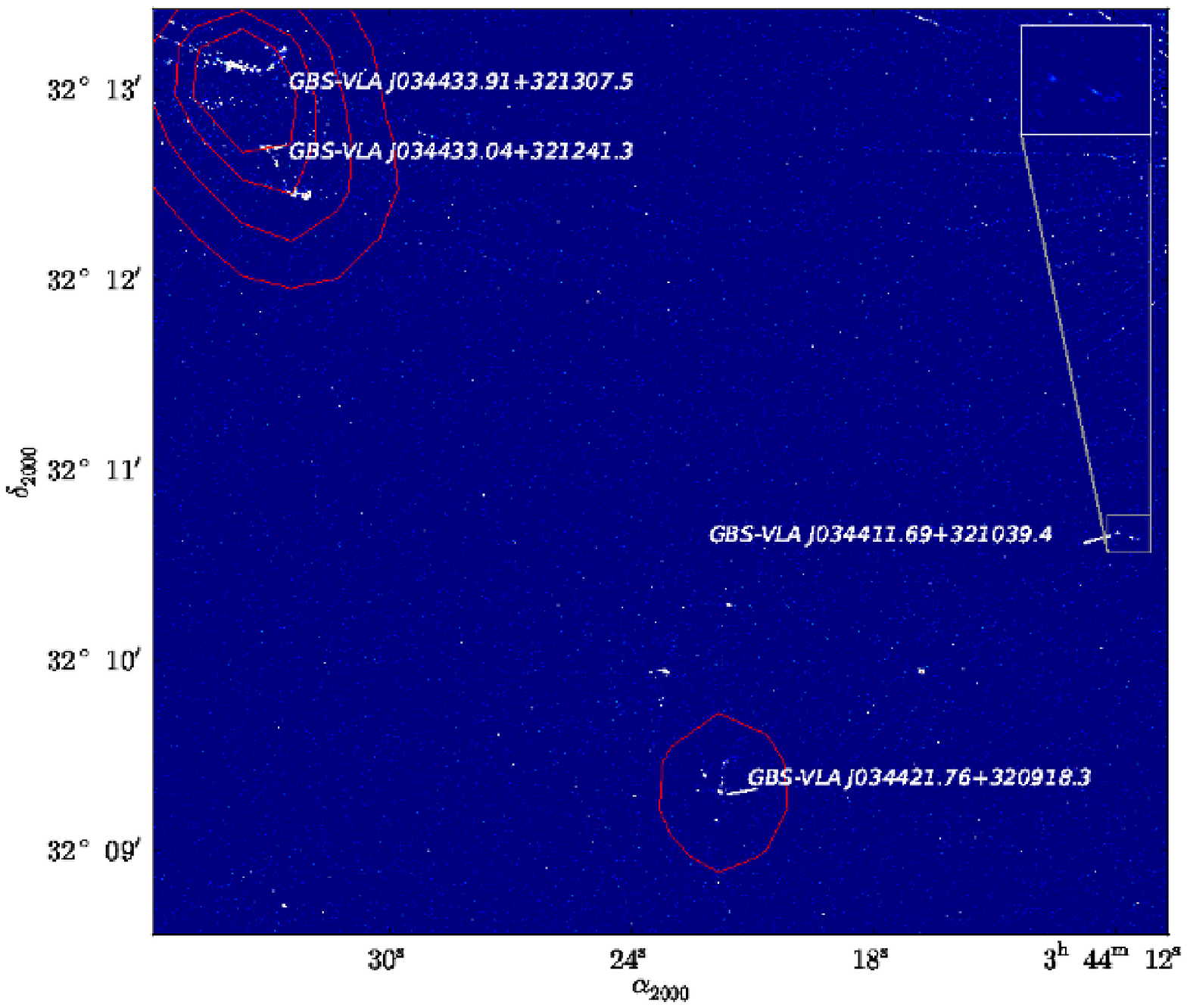}
\caption{Radio map showing the extended sources found in IC 348 observations at 7.5 GHz overlaid with 1.4~GHz NVSS observations (red contours; \citealt{1998AJ....115.1693C}). White contours are 4, 10, 10 and 35 times the rms of the image (rms = 15~$\mu$Jy). }
\label{fig:IC348X}
\end{center}
\end{figure}

\begin{figure}[htbp]
\begin{center}
\includegraphics{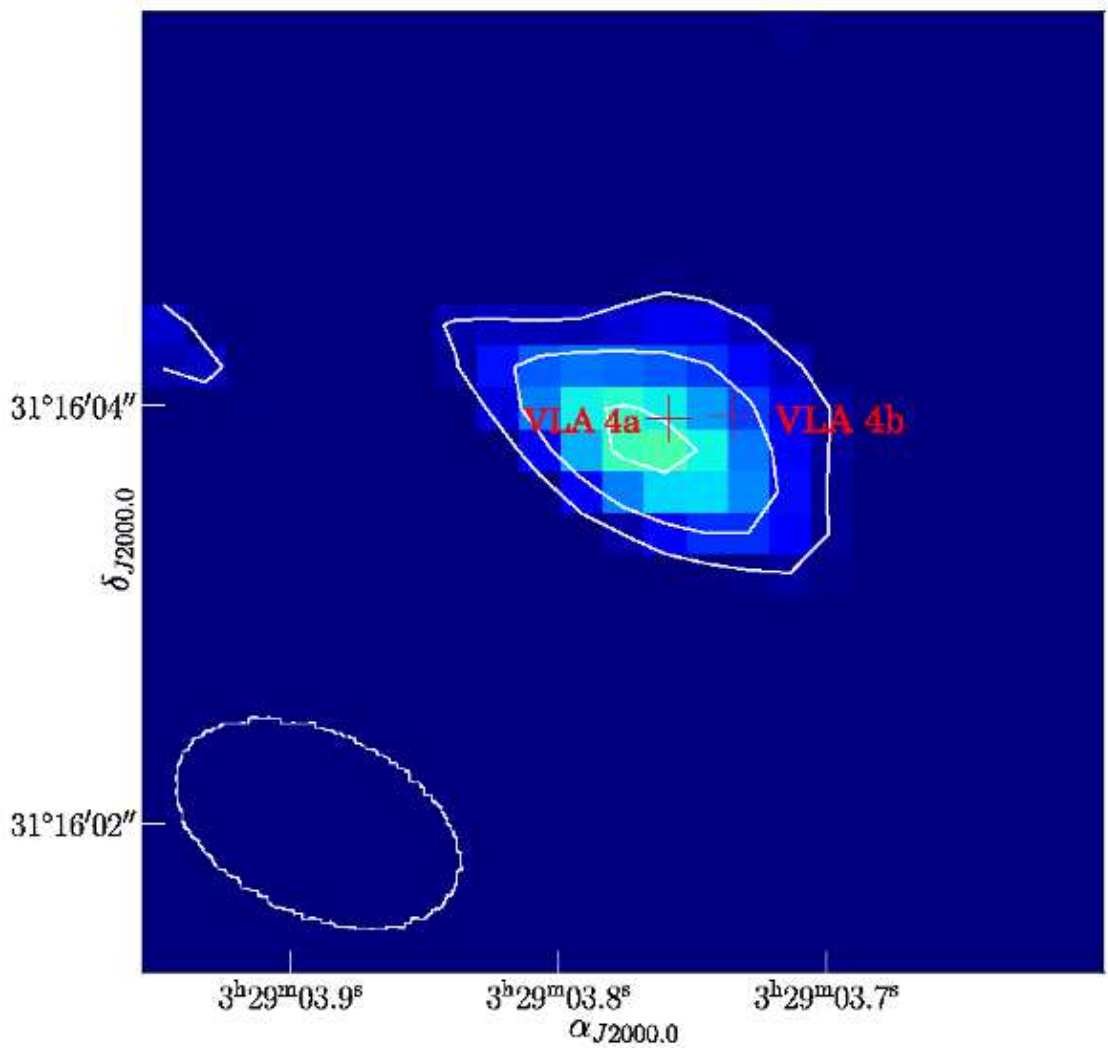}
\caption{Radio map of source GBS-VLA J032903.75+311603.7 at 4.5 GHz. Contours are 2, 4 and 7 times the rms of the image (rms = 16~$\mu$Jy). The synthesized beam, shown in the bottom-left corner, is $1\farcs4 \times 0\farcs9$; P. A. = $65\degr$. The red crosses indicates reported positions of VLA 4a and VLA 4b \citep{2004ApJ...605L.137A}.}
\label{fig:VLA4}
\end{center}
\end{figure}

\begin{figure}[htbp]
\begin{center}
\includegraphics{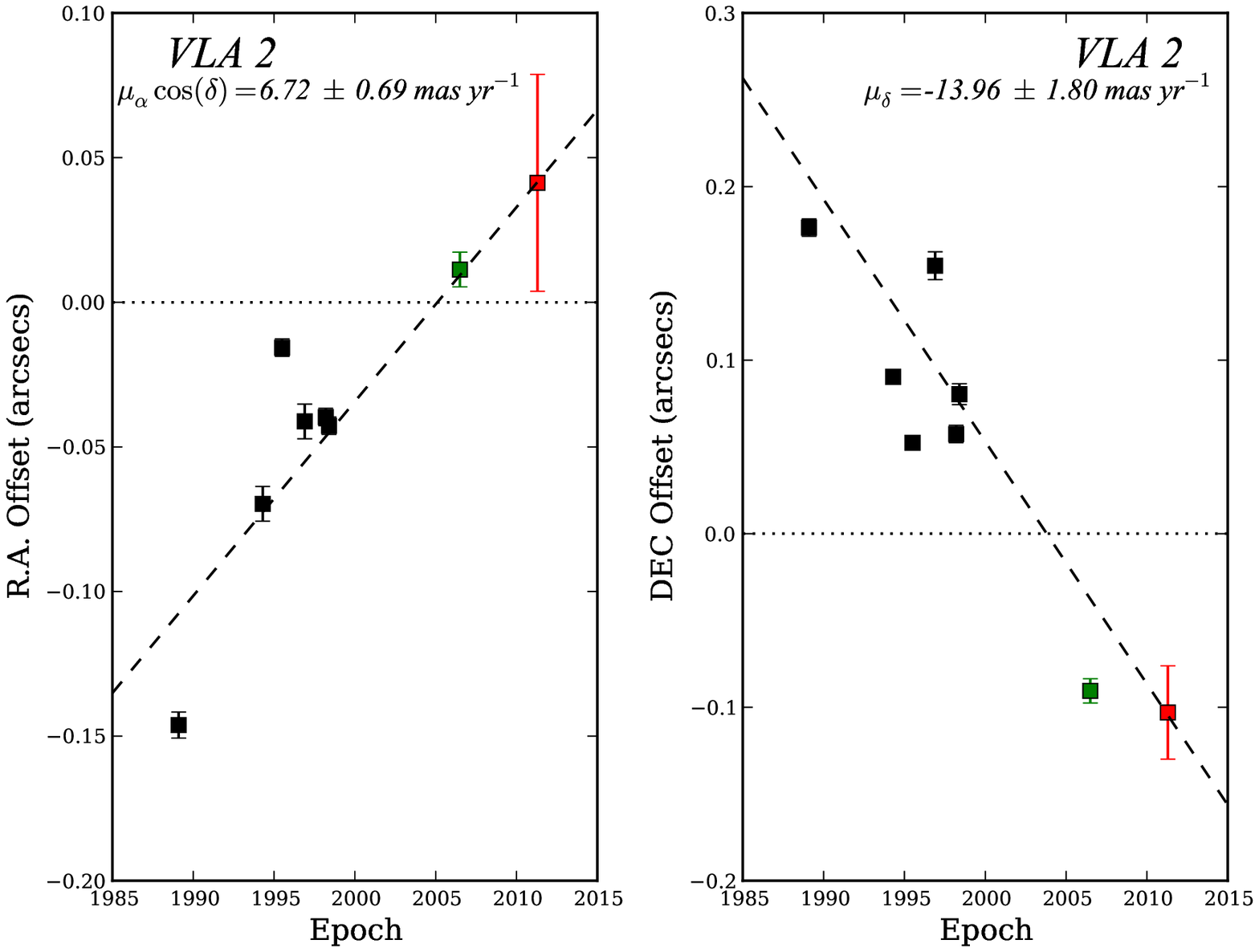}
\caption{Position vs. time diagrams of the source GBS-VLA J032901.97+311538.1 (VLA 2). The position of the source is represented in the diagrams as right ascension and declination offsets (in arcsecs) relative to the mean position.  The black squares are positions reported by \citet{2008AJ....136.2238C}, green squares are positions reported by \citet{2011ApJ...736...25F} and red squares are positions reported in this work. The dashed line in each panel is a linear least-squares fit to the data.  The values of $\mu_\alpha \cos(\delta)$ and $\mu_\delta$ obtained from this fits are also labeled in each panel.}
\label{fig:pmvla2}
\end{center}
\end{figure}

\begin{figure}[htbp]
\begin{center}
\includegraphics{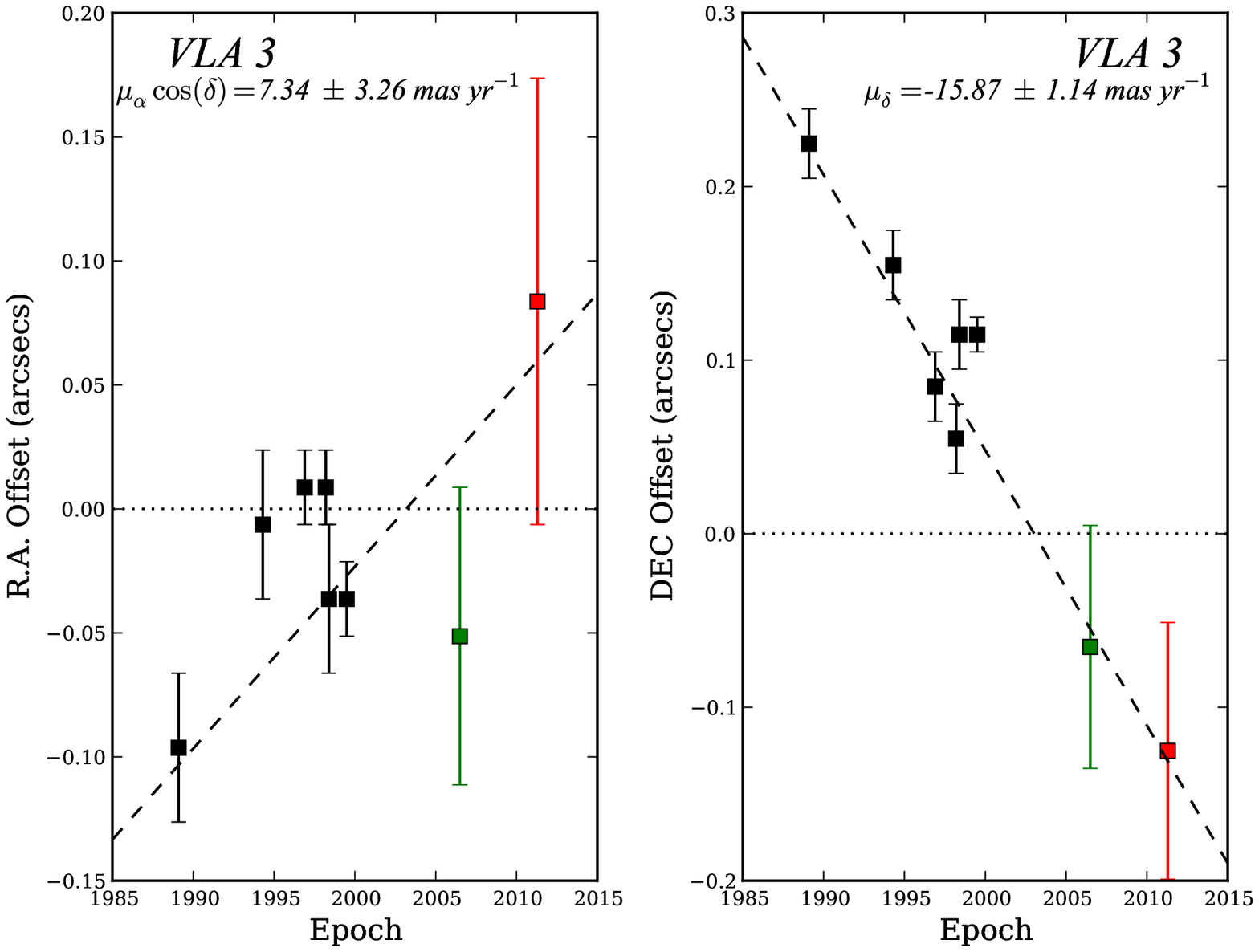}
\caption{Position vs. time diagrams of the source GBS-VLA J032903.38+311601.7 (VLA 3). The position of the source is represented in the diagrams as right ascension and declination offsets (in arcsecs) relative to the mean position.  The black squares are positions reported by \citet{2008AJ....136.2238C}, green squares are positions reported by \citet{2011ApJ...736...25F} and red squares are positions reported in this work. The dashed line in each panel is a linear least-squares fit to the data.  The values of $\mu_\alpha \cos(\delta)$ and $\mu_\delta$ obtained from this fits are also labeled in each panel.}
\label{fig:pmvla3}
\end{center}
\end{figure}

As Figure \ref{fig:map} shows, our observations cover only a fraction of the Perseus complex. Similar large scale observations of other sub-clouds (e.g. B5, B1, L1455 or L1448) would provide a very interesting complement to the observations presented here. 

G.P., L.L., L.F.R., G.N.O., J.L.R. and L.A.Z., acknowledge the financial support of DGAPA, UNAM, and CONACyT,  M\'exico. The National Radio Astronomy Observatory is operated by Associated Universities, Inc., under cooperative agreement with the National Science Foundation. CASA is developed by an international consortium of scientists based at the National Radio Astronomical Observatory (NRAO), the European Southern Observatory (ESO), the National Astronomical Observatory of Japan (NAOJ), the CSIRO Australia Telescope National Facility (CSIRO/ATNF), and the Netherlands Institute for Radio Astronomy (ASTRON) under the guidance of NRAO. This research has made use of the SIMBAD database and VizieR Catalogue Service, operated at CDS, Strasbourg, France

\clearpage

\begin{deluxetable}{lccccc}
\tabletypesize{\scriptsize}
\tablewidth{0pt}
\tablecolumns{6}
\tablecaption{JVLA Observations. \label{tab:resolution}}
\tablehead{Region & Epoch\tablenotemark{a} & \multicolumn{2}{c}{Synthesized Beam\tablenotemark{b}} & \multicolumn{2}{c}{rms Noise\tablenotemark{c}} \\
& & \multicolumn{2}{c}{($\theta_{maj} \times \theta_{min}$; P.A.)} & \multicolumn{2}{c}{($\mu$Jy~beam$^{-1}$)}\\ 
\cmidrule(rl){3-4} \cmidrule(lr){5-6}
& & 4.5 GHz & 7.5 GHz & 4.5 GHz & 7.5GHz}
\startdata
NGC 1333 & 1 & 1\farcs29$\times$1\farcs00, 97.7 & 0\farcs80$\times$0\farcs61, 99.4 & 26 & 23 \\
 & 2 & 2\farcs43$\times$0\farcs94, 75.7 & 1\farcs52$\times$0\farcs58, 74.6 & 35 & 30 \\
 & 3 & 2\farcs26$\times$0\farcs81, 63.5 & 1\farcs32$\times$0\farcs64, 65.9 & 38 & 31 \\
 & C & 1\farcs40$\times$0\farcs90, 65.0 & 0\farcs90$\times$0\farcs50, 66.0 & 16 & 18 \\
 IC 348 & 1 & 1\farcs10$\times$1\farcs01, $-$24.5 & 0\farcs68$\times$0\farcs61, $-$23.0 & 18 & 18 \\
 & 2 & 1\farcs59$\times$1\farcs02, 85.6 & 1\farcs00$\times$0\farcs64, 62.6 & 21 & 25\\
 & 3 & 1\farcs41$\times$0\farcs37, 62.7 & 0\farcs92$\times$0\farcs23, 62.6 & 33 & 27\\
 & C & 1\farcs17$\times$0\farcs63, 65.6 & 0\farcs79$\times$0\farcs34, 62.6 & 16 & 15\\
 Singles\tablenotemark{d} & 1 & 1\farcs37$\times$1\farcs06, 98.7 & 0\farcs84$\times$0\farcs64, 98.9 & 35 & 37\\
 & 2 & 1\farcs64$\times$0\farcs35, 90.9 & 0\farcs99$\times$0\farcs22, 100.17 & 47 & 50 \\
 & 3 & 1\farcs11$\times$0\farcs63, 65.6 & 0\farcs79$\times$0\farcs34, $-$37.0 & 51 & 48\\
 & C & 1.3\farcs$\times$0\farcs54, 99.0 & 0\farcs82$\times$0\farcs32, $-$79.5 & 29 & 28
\enddata
\tablenotetext{a}{C indicates parameters measured in the images after combining the epochs.}
\tablenotetext{b}{Units are arcseconds and degrees respectively.}
\tablenotetext{c}{Measured at the center of the Stokes \textit{I} image.}
\tablenotetext{d}{Average values in the seven individual fields.}
\end{deluxetable}

\clearpage

\begin{deluxetable}{cccrcrc}
\tabletypesize{\scriptsize}
\tablewidth{0pt}
\tablecolumns{7}
\tablecaption{Radio sources detected in NGC1333 \label{tab:sourNGC1333}}
\tablehead{    &  &      \multicolumn{4}{c}{Flux Properties}	&Spectral \\
\colhead{GBS-VLA Name} & \colhead{New Source}\tablenotemark{1} &\colhead{$f_{4.5}$(mJy)} &\colhead{Var.$_{4.5}$\,(\%)} & \colhead{$f_{7.5}$(mJy)} & \colhead{Var.$_{7.5}$\,(\%)}&\colhead{Index}\\}
\startdata 
J032813.80+311755.1 & Y & 0.10 $\pm$ 0.02 & $>$69 $\pm$  11 & -- & -- & -- \\
J032819.46+311831.0 & Y & 0.65 $\pm$ 0.10 &  36 $\pm$  15 & 0.19 $\pm$ 0.06 & $>$23 $\pm$  34 & -2.5 $\pm$  0.3 \\
J032820.31+312509.4 & Y & 0.15 $\pm$ 0.03 &  31 $\pm$  32 & 0.05 $\pm$ 0.03 & $>$16 $\pm$  46 & -2.0 $\pm$  0.7 \\
J032821.37+311440.1 & N & 0.37 $\pm$ 0.06 &  35 $\pm$  18 & -- & -- & -- \\
J032821.70+311555.0 & Y & 0.07 $\pm$ 0.02 & $>$26 $\pm$  28 & $<$0.06 & -- & $<$-0.4 $\pm$  0.5 \\
J032822.25+311427.1 & N & 0.43 $\pm$ 0.07 &  33 $\pm$  19 & -- & -- & -- \\
J032825.98+311616.0 & Y & 0.34 $\pm$ 0.04 &  52 $\pm$  11 & 0.08 $\pm$ 0.02 & $>$18 $\pm$  35 & -3.0 $\pm$  0.5 \\
J032826.24+312440.6 & Y & 0.07 $\pm$ 0.02 &  52 $\pm$  34 & $<$0.07 & -- & $<$-0.1 $\pm$  0.5 \\
J032832.21+313012.7 & Y & 0.08 $\pm$ 0.02 & $>$45 $\pm$  19 & -- & -- & -- \\
J032832.41+311245.3 & Y & 0.12 $\pm$ 0.03 & $>$ 4 $\pm$  33 & -- & -- & -- \\
J032832.87+311445.3 & N & 1.10 $\pm$ 0.12 &  16 $\pm$  14 & 0.59 $\pm$ 0.14 &  31 $\pm$  25 & -1.3 $\pm$  0.2 \\
J032833.25+313043.5 & Y & 0.09 $\pm$ 0.02 & $>$64 $\pm$  21 & -- & -- & -- \\
J032837.01+312125.3 & N & 0.43 $\pm$ 0.03 &  21 $\pm$  10 & 0.35 $\pm$ 0.03 &  30 $\pm$  12 & -0.4 $\pm$  0.2 \\
J032837.10+311330.7 & N & 0.10 $\pm$ 0.02 & $>$25 $\pm$  24 & -- & -- & -- \\
J032843.65+311702.7 & N & 0.08 $\pm$ 0.02 & $>$ 8 $\pm$  43 & -- & -- & -- \\
J032846.49+312943.5 & Y & 0.46 $\pm$ 0.10 & $>$87 $\pm$   4 & -- & -- & -- \\
J032849.45+312841.7 & N & 1.50 $\pm$ 0.22 &  14 $\pm$  18 & 0.90 $\pm$ 0.29 &  56 $\pm$  20 & -1.0 $\pm$  0.2 \\
J032850.72+312225.2 & N & 1.40 $\pm$ 0.09 &  17 $\pm$   8 & 1.27 $\pm$ 0.11 &   5 $\pm$  14 & -0.2 $\pm$  0.2 \\
J032851.99+310924.6 & Y & 0.09 $\pm$ 0.03 & $>$36 $\pm$  23 & -- & -- & -- \\
J032856.92+311622.2 & N & 0.10 $\pm$ 0.02 & $>$38 $\pm$  15 & 0.12 $\pm$ 0.03 & --A-- &  0.5 $\pm$  0.6 \\
J032857.30+311531.4 & N & 0.06 $\pm$ 0.02 & $>$72 $\pm$   9 & -- & -- & -- \\
J032857.36+311415.8 & N & 0.15 $\pm$ 0.02 &  15 $\pm$  27 & 0.18 $\pm$ 0.03 &  20 $\pm$  23 &  0.3 $\pm$  0.3 \\
J032857.37+312954.0 & Y & 0.09 $\pm$ 0.02 &  35 $\pm$  46 & -- & -- & -- \\
J032857.40+312953.9 & Y & 0.10 $\pm$ 0.02 & $>$36 $\pm$  36 & -- & -- & -- \\
J032857.65+311531.4 & N & 0.56 $\pm$ 0.05 &  23 $\pm$  10 & 0.33 $\pm$ 0.05 &  39 $\pm$  19 & -1.1 $\pm$  0.2 \\
J032859.25+312033.0 & N & 0.12 $\pm$ 0.02 &  43 $\pm$  22 & 0.12 $\pm$ 0.03 &  42 $\pm$  26 &  0.1 $\pm$  0.5 \\
J032859.27+311548.2 & N & 0.11 $\pm$ 0.02 & $>$84 $\pm$   2 & 0.09 $\pm$ 0.02 & $>$70 $\pm$   9 & -0.5 $\pm$  0.5 \\
J032859.66+312542.7 & N & 0.12 $\pm$ 0.02 &  35 $\pm$  24 & -- & -- & -- \\
J032859.83+311402.8 & N & 0.16 $\pm$ 0.02 &  24 $\pm$  19 & 0.13 $\pm$ 0.02 & $>$43 $\pm$  11 & -0.4 $\pm$  0.4 \\
J032900.23+313029.7 & Y & 0.09 $\pm$ 0.03 & $>$40 $\pm$  29 & -- & -- & -- \\
J032900.30+312957.6 & Y & 0.07 $\pm$ 0.02 & $>$33 $\pm$  19 & -- & -- & -- \\
J032900.37+312045.4 & N & 0.08 $\pm$ 0.02 & $>$50 $\pm$  10 & 0.05 $\pm$ 0.02 & $>$ 6 $\pm$  31 & -1.0 $\pm$  0.9 \\
J032901.21+312026.0 & N & 0.12 $\pm$ 0.02 &  34 $\pm$  28 & 0.09 $\pm$ 0.03 & --A-- & -0.5 $\pm$  0.6 \\
J032901.63+312018.6 & N & 0.11 $\pm$ 0.02 &  28 $\pm$  26 & 0.08 $\pm$ 0.03 & $>$14 $\pm$  37 & -0.6 $\pm$  0.6 \\
J032901.96+311538.0 & N & 1.03 $\pm$ 0.08 &   5 $\pm$  12 & 1.08 $\pm$ 0.15 &  22 $\pm$  17 &  0.1 $\pm$  0.2 \\
J032902.43+312924.6 & Y & 0.07 $\pm$ 0.02 & $>$32 $\pm$  26 & -- & -- & -- \\
J032903.14+312752.6 & Y & 0.11 $\pm$ 0.02 &  17 $\pm$  39 & $<$0.06 & -- & $<$-1.4 $\pm$  0.3 \\
J032903.38+311601.6 & N & 0.08 $\pm$ 0.02 & $>$51 $\pm$  15 & 0.05 $\pm$ 0.02 & $>$36 $\pm$  28 & -0.8 $\pm$  0.8 \\
J032903.75+311603.7 & N & 0.10 $\pm$ 0.02 &  69 $\pm$  19 & 0.21 $\pm$ 0.04 &  35 $\pm$  28 &  1.5 $\pm$  0.4 \\
J032904.06+311446.2 & N & 0.07 $\pm$ 0.02 & --A-- & $<$0.05 & $>$52 $\pm$  24 & $<$-0.5 $\pm$  0.5 \\
J032904.26+311609.0 & N & 0.12 $\pm$ 0.02 &  63 $\pm$  17 & 0.07 $\pm$ 0.02 &  24 $\pm$  68 & -1.0 $\pm$  0.6 \\
J032907.13+312635.2 & Y & 0.14 $\pm$ 0.02 &  44 $\pm$  21 & 0.10 $\pm$ 0.02 &  15 $\pm$  32 & -0.6 $\pm$  0.5 \\
J032907.16+311708.9 & N & 0.09 $\pm$ 0.02 &  19 $\pm$  30 & $<$0.06 & -- & $<$-0.9 $\pm$  0.4 \\
J032907.75+312157.1 & N & 0.08 $\pm$ 0.02 & $>$28 $\pm$  20 & 0.19 $\pm$ 0.03 &  35 $\pm$  25 &  1.7 $\pm$  0.5 \\
J032907.87+312348.0 & Y & 0.10 $\pm$ 0.02 & $>$57 $\pm$  13 & 0.05 $\pm$ 0.02 & $>$37 $\pm$  29 & -1.4 $\pm$  1.0 \\
J032909.14+312144.0 & N & 0.11 $\pm$ 0.02 & $>$67 $\pm$   7 & 0.15 $\pm$ 0.03 & $>$68 $\pm$  14 &  0.6 $\pm$  0.5 \\
J032909.64+311450.5 & Y & 0.09 $\pm$ 0.02 & $>$39 $\pm$  36 & 0.07 $\pm$ 0.02 & $>$49 $\pm$  32 & -0.4 $\pm$  0.7 \\
J032910.22+312335.1 & N & 0.05 $\pm$ 0.02 & $>$18 $\pm$  30 & $<$0.06 & -- & $<$ 0.3 $\pm$  0.7 \\
J032910.39+312159.0 & N & 0.48 $\pm$ 0.03 &  16 $\pm$  12 & 0.35 $\pm$ 0.03 &   7 $\pm$  23 & -0.6 $\pm$  0.2 \\
J032910.42+311332.0 & N & 0.05 $\pm$ 0.02 & $>$17 $\pm$ 31 & 0.14 $\pm$ 0.03 & $>$38 $\pm$  33 &  2.1 $\pm$  0.7 \\
J032910.53+311330.9 & N & 0.07 $\pm$ 0.02 & $>$ 4 $\pm$  29 & 0.11 $\pm$ 0.03 & $>$36 $\pm$  20 &  1.0 $\pm$  0.6 \\
J032911.25+311831.1 & N & 0.16 $\pm$ 0.02 &  42 $\pm$  13 & 0.19 $\pm$ 0.02 &  41 $\pm$  20 &  0.3 $\pm$  0.3 \\
J032914.11+313057.5 & Y & 0.11 $\pm$ 0.03 & $>$38 $\pm$  24 & -- & -- & -- \\
J032915.85+311621.4 & N & 0.10 $\pm$ 0.02 & $>$52 $\pm$  13 & 0.11 $\pm$ 0.04 & $>$36 $\pm$  22 &  0.2 $\pm$  0.8 \\
J032916.59+311648.7 & N & 0.08 $\pm$ 0.02 & $>$53 $\pm$  15 & 0.14 $\pm$ 0.04 & $>$61 $\pm$  13 &  1.2 $\pm$  0.6 \\
J032917.67+312244.9 & N & 0.09 $\pm$ 0.02 & $>$31 $\pm$  19 & 0.13 $\pm$ 0.03 &  32 $\pm$  41 &  0.8 $\pm$  0.5 \\
J032918.56+311427.3 & Y & 0.07 $\pm$ 0.02 &  76 $\pm$  15 & -- & -- & -- \\
J032920.35+312108.5 & Y & 0.10 $\pm$ 0.02 & $>$34 $\pm$  29 & -- & -- & -- \\
J032920.67+311549.5 & N & 2.10 $\pm$ 0.26 &  Extended & 0.98 $\pm$ 0.24 &  51 Extended & -1.5 $\pm$  0.2 \\
J032922.29+311354.2 & N & 0.30 $\pm$ 0.05 &  21 $\pm$  21 & 0.22 $\pm$ 0.08 &  21 $\pm$  43 & -0.6 $\pm$  0.3 \\
J032923.95+311620.0 & N & 0.49 $\pm$ 0.05 &  16 $\pm$  14 & 0.28 $\pm$ 0.06 &   9 $\pm$  39 & -1.2 $\pm$  0.3 \\
J032926.55+310937.0 & Y & 0.07 $\pm$ 0.02 & $>$58 $\pm$  24 & -- & -- & -- \\
J032926.57+312254.5 & Y & 0.05 $\pm$ 0.02 & $>$55 $\pm$  28 & -- & -- & -- \\
J032927.38+312255.2 & Y & 0.04 $\pm$ 0.02 & $>$15 $\pm$  28 & -- & -- & -- \\
J032929.04+312802.4 & Y & 0.08 $\pm$ 0.02 &  49 $\pm$  28 & $<$0.06 & $>$11 $\pm$  34 & $<$-0.5 $\pm$  0.5 \\
J032930.94+312211.8 & N & 0.85 $\pm$ 0.05 &  16 $\pm$   9 & 0.76 $\pm$ 0.05 &   4 $\pm$  15 & -0.2 $\pm$  0.2 \\
J032931.95+312121.9 & N & 0.06 $\pm$ 0.02 & $>$29 $\pm$  24 & -- & -- & -- \\
J032933.19+312845.2 & Y & 0.07 $\pm$ 0.02 & $>$71 $\pm$  24 & $<$0.07 & -- & $<$-0.2 $\pm$  0.5 \\
J032933.78+311800.8 & Y & 0.05 $\pm$ 0.02 & $>$44 $\pm$  28 & -- & -- & -- \\
J032936.98+311701.9 & Y & 0.09 $\pm$ 0.02 & $>$47 $\pm$  27 & 0.06 $\pm$ 0.02 & $>$28 $\pm$  39 & -0.9 $\pm$  0.8 \\
J032939.39+312309.2 & Y & 0.14 $\pm$ 0.04 & $>$47 $\pm$  27 & $<$0.06 & -- & $<$-1.9 $\pm$  0.6 \\
J032944.99+312019.7 & Y & 0.15 $\pm$ 0.03 & $>$58 $\pm$  18 & $<$0.06 & -- & $<$-1.8 $\pm$  0.4 \\
J032946.15+312353.7 & Y & 0.06 $\pm$ 0.02 & $>$27 $\pm$  30 & -- & -- & -- \\
J032950.32+312646.5 & Y & 0.08 $\pm$ 0.03 & $>$51 $\pm$  27 & -- & -- & -- 
\enddata
\tablecomments{The A annotation indicates a source not detected at three times the noise level on individual epochs, but detected on the image of the concatenated epochs.}
\tablenotetext{1}{Y = source without reported counterparts at any frequency. N = source with known counterpart.}
\end{deluxetable}

\clearpage

\begin{deluxetable}{cccrcrc}
\tabletypesize{\scriptsize}
\tablewidth{0pt}
\tablecolumns{7}
\tablecaption{Radio sources detected in IC 348 \label{tab:sourIC348}}
\tablehead{      &     & \multicolumn{4}{c}{Flux Properties}	&Spectral\\
\colhead{GBS-VLA Name} & \colhead{New Source}\tablenotemark{1} &\colhead{$f_{4.5}$(mJy)} &\colhead{Var.$_{4.5}$(\%)} & \colhead{$f_{7.5}$(mJy)} & \colhead{Var.$_{7.5}$(\%)}&\colhead{Index}\\}
\startdata 
J034311.03+320226.4 & Y & 0.13 $\pm$ 0.03 &  61 $\pm$  22 & -- & -- & -- \\
J034313.02+320242.3 & Y & 0.08 $\pm$ 0.02 & $>$47 $\pm$  25 & -- & -- & -- \\
J034314.84+320947.6 & Y & 0.11 $\pm$ 0.04 & $>$38 $\pm$  27 & -- & -- & -- \\
J034327.28+320028.1 & Y & 0.14 $\pm$ 0.02 & $>$28 $\pm$  17 & 0.13 $\pm$ 0.03 & $>$43 $\pm$  16 & -0.2 $\pm$  0.4 \\
J034330.40+320758.4 & Y & 0.22 $\pm$ 0.02 &  41 $\pm$  14 & -- & -- & -- \\
J034331.68+321451.9 & N & 6.19 $\pm$ 1.08 &  21 $\pm$  20 & -- & -- & -- \\
J034333.93+321307.4 & Y & 10.98 $\pm$ 1.90 &  33 $\pm$  16 & -- & -- & -- \\
J034334.66+320721.1 & Y & 0.13 $\pm$ 0.03 &   8 $\pm$  68 & $<$0.05 & -- & $<$-2.0 $\pm$  0.4 \\
J034337.86+320649.2 & N & 0.05 $\pm$ 0.02 & $>$60 $\pm$  21 & $<$0.05 & -- & $<$-0.3 $\pm$  0.6 \\
J034341.27+315754.1 & N & 0.07 $\pm$ 0.02 & $>$47 $\pm$  25 & -- & -- & -- \\
J034342.10+320225.2 & Y & 0.12 $\pm$ 0.02 & $>$40 $\pm$  13 & -- & -- & -- \\
J034346.30+321039.7 & Y & 0.33 $\pm$ 0.03 &  17 $\pm$  16 & 0.14 $\pm$ 0.02 & $>$49 $\pm$  11 & -1.8 $\pm$  0.3 \\
J034346.85+321814.8 & Y & 0.35 $\pm$ 0.06 &  41 $\pm$  16 & -- & -- & -- \\
J034347.85+320555.2 & Y & 0.08 $\pm$ 0.02 & $>$63 $\pm$  21 & -- & -- & -- \\
J034347.95+315743.6 & Y & 0.09 $\pm$ 0.03 & $>$27 $\pm$  13 & -- & -- & -- \\
J034351.23+321309.1 & N & 0.12 $\pm$ 0.02 &  51 $\pm$  13 & 0.08 $\pm$ 0.02 & $>$49 $\pm$  19 & -0.8 $\pm$  0.5 \\
J034355.23+320057.2 & Y & 0.10 $\pm$ 0.04 & $>$48 $\pm$  12 & -- & -- & -- \\
J034355.41+321008.7 & N & 0.07 $\pm$ 0.02 & $>$46 $\pm$  21 & -- & -- & -- \\
J034355.52+320924.4 & Y & 0.06 $\pm$ 0.02 & $>$55 $\pm$  20 & -- & -- & -- \\
J034356.01+320928.4 & N & 0.21 $\pm$ 0.03 &  27 $\pm$  26 & 0.11 $\pm$ 0.02 &  66 $\pm$  28 & -1.4 $\pm$  0.4 \\
J034356.39+321042.6 & N & 0.42 $\pm$ 0.04 &  29 $\pm$  10 & 0.28 $\pm$ 0.04 &  13 $\pm$  23 & -0.8 $\pm$  0.2 \\
J034357.60+320137.3 & N & 0.21 $\pm$ 0.02 &  76 $\pm$   6 & 0.17 $\pm$ 0.02 & $>$76 $\pm$   7 & -0.4 $\pm$  0.3 \\
J034358.35+315754.7 & Y & 0.08 $\pm$ 0.02 & $>$66 $\pm$  21 & -- & -- & -- \\
J034359.65+320153.9 & N & 0.12 $\pm$ 0.02 & $>$76 $\pm$   4 & 0.14 $\pm$ 0.03 & $>$79 $\pm$   7 &  0.4 $\pm$  0.4 \\
J034401.19+321230.4 & N & 0.26 $\pm$ 0.03 &  25 $\pm$  15 & 0.14 $\pm$ 0.03 &  21 $\pm$  34 & -1.3 $\pm$  0.3 \\
J034401.57+321232.4 & Y & 0.14 $\pm$ 0.02 &  41 $\pm$  18 & 0.08 $\pm$ 0.02 & $>$29 $\pm$  19 & -1.1 $\pm$  0.6 \\
J034402.03+315813.9 & Y & 0.07 $\pm$ 0.02 & $>$27 $\pm$  38 & -- & -- & -- \\
J034404.17+321526.2 & Y & 0.16 $\pm$ 0.02 & $>$49 $\pm$   9 & $<$0.05 & -- & $<$-2.5 $\pm$  0.3 \\
J034405.59+321938.8 & Y & 0.15 $\pm$ 0.02 &  17 $\pm$  23 & 0.10 $\pm$ 0.02 & $>$ 2 $\pm$  32 & -0.8 $\pm$  0.4 \\
J034406.38+321409.8 & N & 0.25 $\pm$ 0.03 &  25 $\pm$  16 & 0.18 $\pm$ 0.02 &  64 $\pm$  24 & -0.6 $\pm$  0.3 \\
J034406.65+321236.9 & N & 4.56 $\pm$ 0.32 &   6 $\pm$   9 & 3.44 $\pm$ 0.41 &  11 $\pm$  15 & -0.6 $\pm$  0.1 \\
J034408.85+320614.1 & N & 0.04 $\pm$ 0.02 &  51 $\pm$  46 & -- & -- & -- \\
J034411.14+320314.0 & N & 0.04 $\pm$ 0.02 & $>$44 $\pm$  39 & $<$0.05 & $>$51 $\pm$  42 & $<$ 0.2 $\pm$  0.8 \\
J034411.69+321039.4 & Y & 2.48 $\pm$ 0.97 & Extended & 0.39 $\pm$ 0.05 & Extended & -- \\
J034412.75+321544.4 & Y & 0.37 $\pm$ 0.04 &  26 $\pm$  12 & 0.41 $\pm$ 0.07 &  35 $\pm$  18 &  0.2 $\pm$  0.2 \\
J034416.02+320513.9 & Y & 0.18 $\pm$ 0.02 &  49 $\pm$  14 & -- & -- & -- \\
J034416.17+321345.5 & N & 0.28 $\pm$ 0.05 &  47 $\pm$  33 & 0.18 $\pm$ 0.03 &  28 $\pm$  23 & -0.8 $\pm$  0.4 \\
J034416.78+320956.4 & N & 2.89 $\pm$ 0.24 &   8 $\pm$  11 & 1.55 $\pm$ 0.24 &  20 $\pm$  18 & -1.3 $\pm$  0.1 \\
J034420.37+320158.4 & N & 1.20 $\pm$ 0.10 &  14 $\pm$  10 & 1.11 $\pm$ 0.16 &   7 $\pm$  20 & -0.1 $\pm$  0.2 \\
J034421.56+321017.4 & N & 0.28 $\pm$ 0.04 & $>$89 $\pm$   1 & 0.35 $\pm$ 0.06 & $>$87 $\pm$   6 &  0.4 $\pm$  0.3 \\
J034421.67+320624.8 & N & 0.08 $\pm$ 0.02 & $>$77 $\pm$   4 & -- & -- & -- \\
J034421.76+320918.3 & N & 3.99 $\pm$ 0.73 & Extended & 2.20 $\pm$ 1.10 &  Extended & -- \\
J034423.11+320956.3 & Y & 1.54 $\pm$ 0.10 & 32 $\pm$ 7 & 0.96 $\pm$ 0.10 & 33 $\pm$ 10 & -1.0 $\pm$  0.2 \\
J034424.57+320357.5 & N & 0.16 $\pm$ 0.03 & $>$59 $\pm$  17 & $<$0.05 & $>$80 $\pm$   8 & $<$-2.4 $\pm$  0.3 \\
J034426.15+320113.1 & N & 0.07 $\pm$ 0.02 &  53 $\pm$  41 & -- & -- & -- \\
J034426.95+315920.0 & Y & 0.07 $\pm$ 0.03 &  84 $\pm$  14 & -- & -- & -- \\
J034427.03+320443.5 & N & 0.07 $\pm$ 0.02 &  87 $\pm$   8 & -- & -- & -- \\
J034431.12+320206.2 & Y & 0.04 $\pm$ 0.02 & $>$51 $\pm$  23 & -- & -- & -- \\
J034431.49+320039.6 & N & 0.22 $\pm$ 0.03 &  29 $\pm$  22 & -- & -- & -- \\
J034431.68+321451.9 & N & 6.21 $\pm$ 0.45 &  20 $\pm$   8 & -- & -- & -- \\
J034432.60+320842.4 & N & 0.39 $\pm$ 0.04 &  72 $\pm$   7 & 0.44 $\pm$ 0.07 & $>$88 $\pm$   6 &  0.2 $\pm$  0.2 \\
J034432.65+321311.9 & Y & 3.91 $\pm$ 0.44 &  46 $\pm$  10 & $<$0.08 & -- & $<$-7.9 $\pm$  0.2 \\
J034432.77+320837.6 & N & 0.12 $\pm$ 0.02 &  30 $\pm$  30 & 0.09 $\pm$ 0.03 & $>$45 $\pm$  33 & -0.5 $\pm$  0.6 \\
J034432.91+321306.6 & N & 1.16 $\pm$ 0.15 &   5 $\pm$  20 & 0.70 $\pm$ 0.14 & $>$77 $\pm$   8 & -1.0 $\pm$  0.4 \\
J034433.04+321241.3 & N & 5.86 $\pm$ 0.36 &  Extended & 4.69 $\pm$ 0.06 &  Extended &  -- \\
J034433.65+321306.4 & N & 3.00 $\pm$ 0.27 &  22 $\pm$  10 & 3.04 $\pm$ 0.44 &  27 $\pm$  16 &  0.0 $\pm$  0.2 \\
J034433.91+321307.5 & N & 12.80 $\pm$ 0.046 &  Extended & 10.03 $\pm$ 0.99 & Extended & -- \\
J034434.05+320104.3 & Y & 0.10 $\pm$ 0.03 & $>$82 $\pm$  14 & -- & -- & -- \\
J034434.87+320633.5 & N & 0.11 $\pm$ 0.02 & $>$42 $\pm$  11 & 0.05 $\pm$ 0.02 & $>$32 $\pm$  39 & -1.6 $\pm$  0.7 \\
J034435.89+320858.7 & N & 0.09 $\pm$ 0.02 & $>$10 $\pm$  24 & $<$0.04 & -- & $<$-1.5 $\pm$  0.4 \\
J034436.47+320313.4 & N & 0.65 $\pm$ 0.11 &  56 $\pm$  12 & 0.10 $\pm$ 0.02 & $>$35 $\pm$  23 & -3.8 $\pm$  0.4 \\
J034436.92+320123.1 & N & 1.29 $\pm$ 0.12 &  23 $\pm$  10 & 0.95 $\pm$ 0.17 &  25 $\pm$  20 & -0.6 $\pm$  0.2 \\
J034436.93+320645.4 & N & 0.34 $\pm$ 0.04 & $>$75 $\pm$   4 & 0.26 $\pm$ 0.06 & $>$80 $\pm$   8 & -0.5 $\pm$  0.3 \\
J034437.73+321839.3 & Y & 0.31 $\pm$ 0.03 & $>$76 $\pm$   4 & -- & -- & -- \\
J034438.48+320820.4 & Y & 0.16 $\pm$ 0.02 & $>$52 $\pm$   9 & 0.09 $\pm$ 0.03 & $>$38 $\pm$  26 & -1.1 $\pm$  0.5 \\
J034438.72+320841.9 & N & 0.18 $\pm$ 0.03 &  61 $\pm$  13 & 0.17 $\pm$ 0.04 & $>$61 $\pm$  12 & -0.2 $\pm$  0.4 \\
J034439.17+320918.4 & N & 0.18 $\pm$ 0.03 &  57 $\pm$  11 & 0.19 $\pm$ 0.04 &  70 $\pm$  10 &  0.2 $\pm$  0.4 \\
J034439.42+320128.8 & Y & 0.09 $\pm$ 0.02 & $>$71 $\pm$  17 & -- & -- & -- \\
J034443.98+320135.2 & N & 0.22 $\pm$ 0.02 &  40 $\pm$  14 & 0.23 $\pm$ 0.03 &  33 $\pm$  23 &  0.1 $\pm$  0.3 \\
J034446.82+320446.5 & N & 0.07 $\pm$ 0.02 & $>$37 $\pm$  25 & 0.06 $\pm$ 0.02 & $>$43 $\pm$  31 & -0.5 $\pm$  0.9 \\
J034446.97+321455.6 & N & 0.93 $\pm$ 0.10 &  52 $\pm$   7 & 0.60 $\pm$ 0.07 &  34 $\pm$  11 & -0.9 $\pm$  0.2 \\
J034447.02+321457.9 & N & 0.54 $\pm$ 0.06 &  43 $\pm$  10 & 0.28 $\pm$ 0.06 & $>$61 $\pm$   9 & -1.3 $\pm$  0.4 \\
J034448.89+320125.0 & Y & 0.12 $\pm$ 0.02 &   6 $\pm$  30 & 0.08 $\pm$ 0.02 & $>$43 $\pm$  45 & -1.0 $\pm$  0.5 \\
J034449.78+315741.7 & Y & 0.08 $\pm$ 0.02 & $>$18 $\pm$  25 & -- & -- & -- \\
J034450.64+321906.3 & N & 0.15 $\pm$ 0.02 & $>$64 $\pm$   9 & 0.14 $\pm$ 0.03 & $>$47 $\pm$  20 & -0.2 $\pm$  0.4 \\
J034452.97+320507.5 & N & 0.47 $\pm$ 0.04 &  26 $\pm$  11 & 0.61 $\pm$ 0.10 &  28 $\pm$  18 &  0.5 $\pm$  0.2 \\
J034453.84+320436.0 & N & 0.33 $\pm$ 0.04 &  25 $\pm$  15 & 0.17 $\pm$ 0.04 &  26 $\pm$  28 & -1.4 $\pm$  0.3 \\
J034458.57+320715.1 & N & 0.04 $\pm$ 0.02 & $>$56 $\pm$  24 & -- & -- & -- \\
J034458.67+315645.8 & Y & 0.06 $\pm$ 0.02 & $>$35 $\pm$  32 & -- & -- & -- \\
J034459.29+315658.9 & Y & 0.37 $\pm$ 0.08 &  46 $\pm$  18 & -- & -- & -- \\
J034507.74+320027.1 & N & 0.06 $\pm$ 0.02 & $>$53 $\pm$  22 & 0.03 $\pm$ 0.02 & $>$58 $\pm$  49 & -1.5 $\pm$  1.3 \\
J034507.97+320401.6 & N & 0.25 $\pm$ 0.02 &  37 $\pm$  10 & 0.24 $\pm$ 0.03 &  51 $\pm$  13 & -0.1 $\pm$  0.2 \\
J034510.90+320822.0 & N & 0.36 $\pm$ 0.04 &  13 $\pm$  16 & 0.38 $\pm$ 0.07 &  13 $\pm$  25 &  0.1 $\pm$  0.2 \\
J034511.72+320219.4 & Y & 0.07 $\pm$ 0.02 & $>$16 $\pm$  28 & $<$0.04 & -- & $<$-1.3 $\pm$  0.5 \\
J034513.19+321001.9 & Y & 0.16 $\pm$ 0.02 &  30 $\pm$  20 & 0.07 $\pm$ 0.02 &  51 $\pm$  47 & -1.7 $\pm$  0.5 \\
J034515.99+320859.7 & Y & 0.29 $\pm$ 0.03 &  41 $\pm$  13 & 0.12 $\pm$ 0.03 &  68 $\pm$  26 & -1.8 $\pm$  0.3 \\
J034516.04+320513.9 & N & 0.18 $\pm$ 0.02 &  33 $\pm$  19 & 0.16 $\pm$ 0.03 &  51 $\pm$  17 & -0.3 $\pm$  0.3 \\
J034519.47+320346.8 & Y & 0.35 $\pm$ 0.04 &  19 $\pm$  18 & 0.20 $\pm$ 0.05 &  18 $\pm$  32 & -1.1 $\pm$  0.2 \\
J034531.69+320400.7 & Y & 0.36 $\pm$ 0.03 &  12 $\pm$  14 & 0.31 $\pm$ 0.04 &  25 $\pm$  17 & -0.3 $\pm$  0.2 \\
J034532.53+320636.9 & N & 1.09 $\pm$ 0.13 &  17 $\pm$  14 & 0.56 $\pm$ 0.13 &  29 $\pm$  26 & -1.4 $\pm$  0.2 \\
J034535.64+320343.5 & Y & 3.13 $\pm$ 0.23 &   6 $\pm$  10 & 1.81 $\pm$ 0.24 &  20 $\pm$  15 & -1.1 $\pm$  0.2 

\enddata
\tablenotetext{1}{Y = source without reported counterparts at any frequency. N = source with known counterpart.}
\end{deluxetable}

\clearpage

\begin{deluxetable}{cccrcrc}
\tabletypesize{\scriptsize}
\tablewidth{0pt}
\tablecolumns{7}
\tablecaption{Radio sources detected in single fields in Perseus \label{tab:soursingles}}
\tablehead{      &     & \multicolumn{4}{c}{Flux Properties}	&Spectral\\
\colhead{GBS-VLA Name} & \colhead{New Source} &\colhead{$f_{4.5}$(mJy)} &\colhead{Var.$_{4.5}$(\%)} & \colhead{$f_{7.5}$(mJy)} & \colhead{Var.$_{7.5}$(\%)}&\colhead{Index}\\}
\startdata 
J032528.40+311109.2 & N & 0.33 $\pm$ 0.08 &  38 $\pm$  23 & -- & -- & -- \\
J032549.54+311408.8 & Y & 0.29 $\pm$ 0.05 &  41 $\pm$  21 & $<$0.17 & -- & $<$-1.0 $\pm$  0.3 \\
J032827.62+304909.4 & N & 0.72 $\pm$ 0.21 & $>$94 $\pm$   2 & -- & -- & -- \\
J032836.79+305017.9 & N & 1.11 $\pm$ 0.27 &  23 $\pm$  27 & -- & -- & -- \\
J032838.21+304007.9 & N & 0.42 $\pm$ 0.10 &  28 $\pm$  28 & -- & -- & -- \\
J032840.88+304948.3 & N & 3.96 $\pm$ 0.86 &  49 $\pm$  16 & -- & -- & -- \\
J032841.15+304945.2 & N & 3.28 $\pm$ 0.73 & $>$92 $\pm$   2 & -- & -- & -- \\
J032852.32+304216.8 & N & 1.60 $\pm$ 0.18 &  32 $\pm$  11 & 1.56 $\pm$ 0.35 &  49 $\pm$  17 & -0.0 $\pm$  0.2 \\
J032855.81+304719.7 & N & 7.05 $\pm$ 0.70 &  10 $\pm$  12 & 4.01 $\pm$ 0.79 &  22 $\pm$  22 & -1.1 $\pm$  0.2 \\
J032906.33+304332.7 & N & 0.40 $\pm$ 0.06 &  34 $\pm$  18 & 0.42 $\pm$ 0.14 &  42 $\pm$  29 &  0.1 $\pm$  0.4 \\
J032912.84+304558.5 & N & 0.45 $\pm$ 0.09 & $>$65 $\pm$  10 & -- & -- & -- \\
J032917.16+304329.7 & N & 3.30 $\pm$ 0.76 &  32 $\pm$  22 & -- & -- & -- \\
J032919.25+304548.7 & Y & 0.59 $\pm$ 0.16 & $>$96 $\pm$   2 & -- & -- & -- \\
J033057.00+313402.9 & Y & 0.33 $\pm$ 0.08 & $>$31 $\pm$  19 & -- & -- & -- \\
J033100.54+313405.7 & N & 0.22 $\pm$ 0.06 & $>$26 $\pm$  25 & -- & -- & -- \\
J033100.76+313412.3 & Y & 0.51 $\pm$ 0.11 &  59 $\pm$  15 & -- & -- & -- \\
J033111.09+313904.7 & N & 2.89 $\pm$ 0.81 &  44 $\pm$  22 & -- & -- & -- \\
J033443.24+315912.8 & Y & 0.19 $\pm$ 0.04 & $>$35 $\pm$  26 & $<$0.21 & -- & $<$ 0.2 $\pm$  0.3 \\
J033454.98+320506.2 & Y & 0.99 $\pm$ 0.16 &  41 $\pm$  14 & 0.79 $\pm$ 0.27 &  68 $\pm$  17 & -0.5 $\pm$  0.2 \\
J033456.97+315806.3 & Y & 0.15 $\pm$ 0.03 & $>$ 7 $\pm$  28 & $<$0.11 & -- & $<$-0.6 $\pm$  0.4 \\
J033501.24+320059.9 & N & 0.34 $\pm$ 0.03 &  63 $\pm$   9 & 0.34 $\pm$ 0.03 & $>$60 $\pm$   8 &  0.0 $\pm$  0.3 \\
J033501.53+320406.0 & Y & 0.42 $\pm$ 0.06 &   4 $\pm$  23 & 0.37 $\pm$ 0.10 &  40 $\pm$  24 & -0.3 $\pm$  0.3 \\
J033503.90+315923.1 & Y & 0.14 $\pm$ 0.03 & $>$34 $\pm$  16 & 0.20 $\pm$ 0.04 & $>$10 $\pm$  28 &  0.7 $\pm$  0.5 \\
J033508.31+315803.3 & N & 0.29 $\pm$ 0.06 &  49 $\pm$  23 & 0.24 $\pm$ 0.08 & --A-- & -0.4 $\pm$  0.5 \\
J033509.29+315802.5 & N & 0.23 $\pm$ 0.05 &  20 $\pm$  28 & 0.11 $\pm$ 0.04 & --A-- & -1.5 $\pm$  0.6 \\
J033517.06+311640.5 & Y & 0.50 $\pm$ 0.09 &  30 $\pm$  19 & 0.13 $\pm$ 0.05 & $>$28 $\pm$  37 & -2.7 $\pm$  0.5 \\
J033517.65+311650.0 & Y & 0.50 $\pm$ 0.08 &  50 $\pm$  15 & 0.83 $\pm$ 0.56 & $>$28 $\pm$  38 &  1.0 $\pm$  1.2 \\
J033530.50+310955.9 & Y & 0.46 $\pm$ 0.07 &  18 $\pm$  21 & 0.25 $\pm$ 0.08 & $>$25 $\pm$  34 & -1.2 $\pm$  0.3 \\
J033541.20+311500.1 & Y & 0.57 $\pm$ 0.07 &  23 $\pm$  16 & 0.41 $\pm$ 0.11 &  44 $\pm$  24 & -0.7 $\pm$  0.2 \\
J033542.19+311727.3 & Y & 0.46 $\pm$ 0.09 &  17 $\pm$  26 & -- & -- & -- \\
J033553.36+310955.7 & Y & 1.46 $\pm$ 0.37 &  38 $\pm$  22 & -- & -- & -- \\
J033620.81+311605.1 & Y & 0.56 $\pm$ 0.09 &  12 $\pm$  21 & 0.57 $\pm$ 0.19 &  25 $\pm$  35 &  0.0 $\pm$  0.3 \\
J033641.02+311753.1 & Y & 1.22 $\pm$ 0.36 &  51 $\pm$  20 & -- & -- & -- \\
J034300.22+293317.1 & N & 6.63 $\pm$ 1.74 &  45 $\pm$  20 & -- & -- & -- \\
J034300.32+293320.4 & N & 2.64 $\pm$ 0.70 &  53 $\pm$  18 & -- & -- & -- \\
J034307.27+293235.0 & N & 1.34 $\pm$ 0.28 &  42 $\pm$  17 & -- & -- & -- \\
J034324.29+293811.5 & Y & 0.43 $\pm$ 0.05 &  19 $\pm$  18 & 0.23 $\pm$ 0.05 & $>$36 $\pm$  21 & -1.3 $\pm$  0.3 \\
J034331.61+293534.5 & N & 1.58 $\pm$ 0.10 &  12 $\pm$   9 & 1.38 $\pm$ 0.14 &   6 $\pm$  15 & -0.3 $\pm$  0.2 \\
J034341.43+293101.0 & Y & 0.46 $\pm$ 0.11 &  52 $\pm$  19 & -- & -- & -- \\
J034344.06+294008.5 & Y & 0.58 $\pm$ 0.13 &  27 $\pm$  24 & -- & -- & -- \\
J034356.20+293620.3 & N & 0.74 $\pm$ 0.20 &  46 $\pm$  23 & -- & -- & -- 
\enddata
\tablecomments{The A annotation indicates a source not detected at three times the noise level on individual epochs, but detected on the image of the concatenated epochs.}
\tablenotetext{1}{Y = source without reported counterparts at any frequency. N = source with known counterpart.}
\end{deluxetable}

\begin{deluxetable}{lccccccc}
\tabletypesize{\scriptsize}
\tablewidth{0pt}
\tablecolumns{8}
\tablecaption{Radio sources with known counterparts in NGC 1333\label{tab:countNGC1333}}
\tablehead{         & Other         &         & \multicolumn{3}{c}{Infrared\tablenotemark{b}} &       & Object\\
\colhead{GBS-VLA Name} & \colhead{Names} &\colhead{X-ray\tablenotemark{a}} & \colhead{SST} & \colhead{2M}&\colhead{WISE}&
\colhead{Radio\tablenotemark{c}}&\colhead{type\tablenotemark{d}}\\}
\startdata
J032821.37+311440.1&--&XMMU J032821.5+311440&--&--&--&--
&X\\
J032822.25+311427.1 & -- & -- & -- & -- & -- & NVSS 032822+311431 & Rad \\
J032832.87+311445.3&--&--&--&--&--&RAC97 VLA 36
&E\\
J032837.01+312125.3&--&--&--&--&--&RAC97 VLA 37
&E\\
J032837.10+311330.7&2MASS J03283706+3113310&--&Y&Y&Y&--
&YSO\\
J032843.65+311702.7&--&WMW2010 19&--&--&--&--
&X\\
J032849.45+312841.7 & -- & -- & -- & -- & -- & NVSS 032848+312844 & Rad \\
J032850.72+312225.2&--&--&--&--&--&RAC97 VLA 38
&YSO\\
J032856.92+311622.2&2MASS J032856.9+311622&--&Y&Y&Y&--
&YSO\\
J032857.30+311531.4&--&--&--&--&--&SB86 NGC1333 7
&Rad\\
J032857.36+311415.8&SSTc2d J032857.4+311416&--&Y&--&--&FOW2011 8
&YSO\\
J032857.65+311531.4&--&--&--&--&--&RAC97 VLA 1
&E\\
J032859.25+312033.0&2MASS J03285920+312037&--&Y&Y&--&--
&YSO?\\
J032859.27+311548.2&2MASS J03285930+3115485&CXO J032859.2+311548&Y&Y&Y&--
&YSO\\
J032859.66+312542.7&--&WMW2010 120&--&--&--&--
&X\\
J032859.83+311402.8&--&--&--&--&--&RAC97 VLA 12
&E\\
J032900.37+312045.4&2MASS J03290037+3120456&CXO J032900.3+312045&Y&Y&Y&--
&YSO\\
J032901.21+312026.0&2MASS J03290116+3120244&--&Y&Y&--&RAC97 VLA 42
&YSO\\
J032901.63+312018.6&2MASS J03290149+3120208&--&Y&Y&Y&--
&YSO\\
J032901.96+311538.0&HH 7-11 MMS 3&--&--&--&--&RAC97 VLA 2
&YSO\\
J032903.38+311601.6&--&--&--&--&--&RAC97 VLA 3
&YSO\\
J032903.75+311603.7&SVS76 NGC 1333 13A1&--&Y&Y&Y&RAC97 VLA 4a/4b
&YSO\\
J032904.06+311446.2&SSTc2d J032904.1+311447&--&Y&Y&Y&RAC97 VLA 19
&YSO\\
J032904.26+311609.0&2MASS J03290421+3116080&--&Y&Y&--&RAC97 VLA 20
&YSO\\
J032907.16+311708.9&--&--&--&--&--&RAC97 VLA 23
&E\\
J032907.75+312157.1&2MASS J03290773+3121575&WMW2010 82&Y&Y&Y&FOW2011 17
&YSO\\
J032909.14+312144.0&2MASS J03290915+3121445&WMW2010 79&Y&Y&--&--
&YSO\\
J032910.22+312335.1&2MASS J03291046+3123348&--&Y&Y&Y&--
&YSO\\
J032910.39+312159.0&2MASS J03291037+3121591&CXO J032910.3+312159&Y&Y&Y&--
&YSO\\
J032910.42+311332.0&JCC87 IRAS 4A2&--&--&--&--&--
& YSO\\
J032910.53+311330.9&JCC87 IRAS 4A1&--&--&--&--&--
& YSO\\
J032911.25+311831.1&SSTc2d J032911.3+311831&--&Y&--&Y&FOW2011 20
&YSO\\
J032915.85+311621.4&--&--&--&--&--&RAC97 VLA 30
&E\\
J032916.59+311648.7&--&--&--&--&--&RAC97 VLA 31
&YSO\\
J032917.67+312244.9&SVS76 NGC 1333 2&CXO J032917.6+312245&Y&Y&Y&FOW2011 21
&YSO\\
J032920.67+311549.5&--&--&--&--&--&RAC97 VLA 32
&E\\
J032922.29+311354.2&GMM2008 38&--&--&--&--&FOW2011 25
&YSO\\
J032923.95+311620.0&--&--&--&--&--&RAC97 VLA 35
&E\\
J032930.94+312211.8&--&--&--&--&--&FOW2011 26
&Rad\\
J032931.95+312121.9&--&CXO J032931.6+312125&--&--&--&
&X\\
\enddata
\tablenotetext{a}{XMMU = \citealt{2003A&A...401..543P}, \citealt{2002A&A...382..522B}; CXO = \citealt{2002ApJ...575..354G}; WMW = \citealt{2010AJ....140..266W}}
\tablenotetext{b}{SST = \citealt{2009ApJS..184...18G}; 2M = \citealt{2003tmc..book.....C} and WISE = \citealt{2012wise.rept....1C}}
\tablenotetext{c}{
RAC97 = \citealt{1997ApJ...480L.125R}; \citealt{1999ApJS..125..427R}; \citealt{2000ApJ...542L.123A}; FOW2011 = \citealt{2011ApJ...736...25F}; NVSS = \citealt{1998AJ....115.1693C}}
\tablenotetext{d}{Object type: E = extragalactic; IR = infrared; Rad = radio source; X = X ray source; YSO = young stellar object; YSO? = young stellar object candidate. Objects are marked as peculiar emitters (Rad, X-ray, IR or smm) when information is not sufficient to determine the nature of the object.} 
\end{deluxetable}

\begin{deluxetable}{lccccccc}
\tabletypesize{\scriptsize}
\tablewidth{0pt}
\tablecolumns{8}
\tablecaption{Radio sources with known counterparts in IC 348 \label{tab:countIC348}}
\tablehead{           & Other         &         & \multicolumn{3}{c}{Infrared\tablenotemark{b}} &       & Object\\
\colhead{GBS-VLA Name} & \colhead{Names} &\colhead{X-ray\tablenotemark{a}} & \colhead{SST} & \colhead{2M}&\colhead{WISE}&
\colhead{Radio\tablenotemark{c}}&\colhead{type}\\}
\startdata
J034331.68+321451.9 & -- & -- & -- & -- & -- & NVSS 034331+321451 & Rad \\
J034337.86+320649.2&JCMTSF J034337.8+320644&--&--&--&--&--
&SMM\\
J034341.27+315754.1&JMCTSE J034341.5+315753&--&--&--&--&--
&SMM\\
J034351.23+321309.1&2MASS J03435123+3213091&CXOPZ 3&Y&Y&Y&--
&YSO\\
J034355.41+321008.7&2MASS J03435519+3210067&--&Y&Y&Y&--
&*\\
J034356.01+320928.4&PSZ2003 J034356.0+320928&--&--&--&--&--
&*\\
J034356.39+321042.6&PSZ2003 J034356.7+321039&--&--&--&--&--
&*\\
J034357.60+320137.3&2MASS J03435759+3201373&CXOU J034357.6+320137&Y&Y&Y&FOW2011 3
&YSO\\
J034359.65+320153.9&2MASS J03435964+3201539&CXOU J034359.6+320154&Y&Y&Y&FOW2011 4
&YSO\\
J034401.19+321230.4&PSZ2003 J034401.2+321230&--&--&--&--&--
&*\\
J034406.38+321409.8&PSZ2003 J034406.4+321410&--&--&--&--&--
&*\\
J034406.65+321236.9 & -- & -- & -- & -- & -- & NVSS 034406+321235 & Rad \\
J034408.85+320614.1&PSZ2003 J034409.2+320613&--&--&--&--&--
&*\\
J034411.14+320314.0&PSZ2003 J034411.0+320315&--&--&--&--&--
&*\\
J034416.17+321345.5&PSZ2003 J034416.2+321345&--&--&--&--&--
&*\\
J034416.78+320956.4&2MASS J03441642+3209552&CXOPZ 32&Y&Y&Y&--
&YSO\\
J034420.37+320158.4&--&CXOPZ 45&--&--&--&FOW2011 6
&YSO\\
J034421.56+321017.4&2MASS J03442155+3210174&CXOPZ 49&Y&Y&Y&--
&YSO\\
J034421.67+320624.8&2MASS J03442166+3206248&CXOPZ 52&Y&Y&Y&--
&YSO\\
J034421.76+320918.3&Cl*IC 348 MM 42&--&--&--&--& NVSS 034421+320918
&*\\
J034424.57+320357.5&2MASS J03442457+3203571&CXOPZ 64&Y&Y&Y&--
&YSO\\
J034426.15+320113.1&--&CXOU J034426.1+320113&--&--&--&--
&X\\
J034427.03+320443.5&2MASS J03442702+3204436&CXOPZ 77&Y&Y&Y&--
&YSO\\
J034431.49+320039.6&--&--&--&--&--&FOW2011 7
&Rad\\
J034431.68+321451.9&PSZ2003 J034431.6+321454&--&--&--&--&--
&*\\
J034432.60+320842.4&2MASS J03443259+3208424&CXOPZ 106&Y&Y&Y&--
&YSO\\
J034432.77+320837.6&2MASS J03443274+3208374&CXOPZ 108&Y&Y&Y&--
&YSO\\
J034432.91+321306.6 & -- & -- & -- & -- & -- & NVSS 034433+321255 & Rad \\
J034433.04+321241.3&--&CXOPZ 110&--&--&--& NVSS 034433+321255
&X,Rad\\
J034433.65+321306.4 & -- & -- & -- & -- & -- & NVSS 034433+321255 & Rad \\
J034433.91+321307.5 & -- & -- & -- & -- & -- & NVSS 034433+321255 & Rad \\
J034434.87+320633.5&2MASS J03443487+3206337&CXOPZ 119&Y&Y&Y&--
&YSO\\
J034435.89+320858.7&Cl*IC 348 MM 149&--&--&--&--&--
&*\\
J034436.47+320313.4&PSZ2003 J034436.5+320317&--&--&--&--&--
&*\\
J034436.92+320123.1&--&CXOPZ 134&--&--&--&--
&Rad\\
J034436.93+320645.4&2MASS J03443694+3206453&CXOPZ 133&Y&Y&Y&--
&YSO\\
J034438.72+320841.9&2MASS J03443871+3208420&CXOPZ 149&Y&Y&Y&--
&YSO\\
J034439.17+320918.4&2MASS J03443916+3209182&CXOPZ 151&Y&Y&Y&--
&YSO\\
J034443.98+320135.2&2MASS J03444389+3201373&--&Y&Y&--&--
&YSO\\
J034446.82+320446.5&PSZ2003 J034446.8+320446&--&--&--&--&--
&*\\
J034446.97+321455.6 & -- & -- & -- & -- & -- & NVSS 034447+321455 & Rad \\
J034447.02+321457.9 & -- & -- & -- & -- & -- & NVSS 034447+321455 & Rad \\
J034450.64+321906.3&2MASS J03445064+3219067&CXOPZ 187&Y&Y&Y&--
&YSO\\
J034452.97+320507.5&PSZ2003 J034453.0+320507&--&--&--&--&--
&*\\
J034453.84+320436.0&PSZ2003 J034453.9+320436&--&--&--&--&--
&*\\
J034458.57+320715.1&PSZ2003 J034458.6+320710&--&--&--&--&--
&*\\
J034507.74+320027.1&2MASS J03450773+3200272&XMMU J034507.6+320027&--&--&--&--
&*\\
J034507.97+320401.6&Cl*IC 348 LRL 11&CXOPZ 209&Y&Y&Y&--
&YSO\\
J034510.90+320822.0&--&CXOPZ 213&--&--&--&--
&X\\
J034516.04+320513.9&2MASS J03451604+3205140&--&Y&Y&Y&--
&* \\
J034532.53+320636.9 & -- & -- &-- & -- &-- & NVSS 034532+320635 & Rad 

\enddata
\tablenotetext{a}{CXOPZ = \citealt{2001AJ....122..866P}; CXOU = Chandra X-ray Observatory, Unregistered; XMMU = \citealt{2001A&A...376..387L}, \citealt{2002A&A...382..522B}}
\tablenotetext{b}{SST = \citealt{2009ApJS..184...18G}; 2M = \citealt{2003tmc..book.....C} and WISE = \citealt{2012wise.rept....1C}}
\tablenotetext{c}{FOW2011 = \citealt{2011ApJ...736...25F}; NVSS = \citealt{1998AJ....115.1693C}}
\tablenotetext{d}{Object type: Rad = radio source; SMM = submillimeter source; X = X-ray source; YSO = young stellar object; * = star; *iC = star in cluster. Objects are marked as peculiar emitters (Rad, X-ray, IR or smm) when information is not sufficient to determine the nature of the object.}
\end{deluxetable}

\begin{deluxetable}{lccccccc}
\tabletypesize{\scriptsize}
\tablewidth{0pt}
\tablecolumns{8}
\tablecaption{Radio sources with known counterparts in Perseus singles fields.\label{tab:countsingles}}
\tablehead{           & Other         &         & \multicolumn{3}{c}{Infrared\tablenotemark{b}} &       & Object\\
\colhead{GBS-VLA Name} & \colhead{Names} &\colhead{X-ray\tablenotemark{a}} & \colhead{SST} & \colhead{2M}&\colhead{WISE}&
\colhead{Radio\tablenotemark{c}}&\colhead{type}\tablenotemark{d}\\}
\startdata
J032528.40+311109.2 & WISE J032528.42+311109.7 & -- & -- & -- & Y & NVSS 032528+311112 & IR, Rad \\
J032827.62+304909.4 & -- & -- & -- &-- & -- & NVSS 032826+304914 & Rad \\
J032836.79+305017.9 & WISE J032836.77+305017.9 & -- & -- & -- & Y & -- & IR \\
J032838.21+304007.9 & WISE J032838.22+304008.0 & -- & -- & -- & Y & -- & IR \\
J032840.88+304948.3&--&--&--&--&--&NVSS J032840+304954
&Rad\\
J032841.15+304945.2 & -- & -- & -- & -- & -- & NVSS 032840+304954 & Rad \\
J032852.32+304216.8 & WISE J032852.32+304217.2 & -- & -- & -- & Y & NVSS 032852+304221 & IR,Rad \\
J032855.81+304719.7 & -- & -- & -- & -- & -- & NVSS 032855+304720 & Rad \\
J032906.33+304332.7 & WISE J032906.29+304333.0 & -- & -- & -- & Y & -- & IR \\
J032912.84+304558.5 & WISE J032912.81+304558.6 & -- & -- & -- & Y & -- & IR \\ 
J032917.16+304329.7 & -- & -- & -- & -- & -- & NVSS 032917+304432 & Rad \\
J033100.54+313405.7 & WISE J033100.51+313405.7 & -- & -- & -- & Y & -- & IR \\
J033111.09+313904.7 & WISE J033111.07+313904.9 & -- & -- & -- & Y & NVSS 033111+313916 & IR, Rad \\
J033501.24+320059.9&V* IX Per&1RXS J033501.2+320104&--&--&--&--
&*\\
J033508.31+315803.3 & WISE J033508.32+315803.1 & -- & -- & -- & Y & -- & IR \\
J033509.29+315802.5 & WISE J033509.30+315803.0 & -- & -- & -- & Y & -- & IR \\
J034300.22+293317.1 & -- & -- & -- & -- & -- & NVSS 034300+293318 & Rad \\
J034300.32+293320.4 & -- & -- & -- & -- & -- & NVSS 034300+293318 & Rad \\
J034307.27+293235.0 & WISE J034307.29+293235.7 & -- & -- & -- & Y & -- & IR \\
J034331.61+293534.5&2MASS J03433201+2935326&--&--&Y&--&--
&* \\
J034356.20+293620.3 & WISE J034356.20+293620.0 & -- & -- & -- & Y & -- & IR 

\enddata
\tablenotetext{a}{1RXS = \citealt{2000IAUC.7432R...1V},\citealt{1999A&A...349..389V}}
\tablenotetext{b}{SST = \citealt{2009ApJS..184...18G}; 2M = \citealt{2003tmc..book.....C} and WISE = \citealt{2012wise.rept....1C}}
\tablenotetext{c}{NVSS = \citealt{1998AJ....115.1693C}}
\tablenotetext{d}{Object type: * = star; IR = infrared source; Rad = radio source. Objects are marked as peculiar emitters (Rad, X-ray, IR or smm) when information is not sufficient to determine the nature of the object.} 
\end{deluxetable}

\clearpage

\begin{deluxetable}{lccccc}
\tabletypesize{\scriptsize}
\tablewidth{0pt}
\tablecolumns{5}
\tablecaption{YSOs with polarized emission. \label{tab:pol}}
\tablehead{GBS-VLA NAME & $F_V/F$\tablenotemark{a} at 4.5 GHz (\%) & $\sigma_{(F_V/F)}$ at 4.5 GHz (\%) & $F_V/F$ at 7.5 GHz (\%) & $\sigma_{(F_V/F)}$ at 7.5 GHz (\%) & Var\tablenotemark{b}}
\startdata
J032850.72+312225.2 & 7.3 & 0.9 & 6.3 & 1.3 & N\\
J032922.29+311354.2 & 37.0 & 4.7 & -- & -- & N\\
J034434.87+320633.5 & 69.6	& 14.5 & -- & -- & N\\
J034450.64+321906.3 & 34.0 & 7.4 & 72.6 & 22.2 & Y 
\enddata
\tablenotetext{a}{F = integrated flux density in the stokes \textit{I} images; F$_\mathrm{V}$ = integrated flux density in the stokes \textit{V} images.}
\tablenotetext{b}{Var.\ = Y when the source variability is higher than 50\% in at least one frequency; N when it is lower.}
\end{deluxetable}

\clearpage

\begin{deluxetable}{lccccccc}
\tabletypesize{\scriptsize}
\tablewidth{0pt}
\tablecolumns{8}
\tablecaption{Young stellar objects Detected\label{tab:YSO}}
\tablehead{GBS-VLA NAME  & Spectral        &      SED\tablenotemark{a}   & Var\tablenotemark{b}&    $\alpha$\tablenotemark{b}   & X-Ray &VLBI \tablenotemark{c} & Ref.\tablenotemark{d}\\
& \colhead{Type} & \colhead{Classification} & & & & Candidates &}
\startdata
\multicolumn{8}{c}{NGC 1333} \\
\hline
J032837.10+311330.7 & -- & Class I & N & -- & N & N &1 \\
J032850.72+312225.2 & -- & -- & N & F & N & Y & 2 \\
J032856.92+311622.2 & M2.5 & Class II & N & P & N & N & 3 \\
J032857.36+311415.8 & -- & Class I & N & P & N & N & 1 \\
J032859.25+312033.0 & -- & Flat & N & F & N & N & 4 \\
J032859.27+311548.2 & K2.0 & Class II & Y & N & Y & N & 3,5 \\
J032900.37+312045.4 & M4.5 & -- & Y & N & Y & N & 3 \\
J032901.21+312026.0 & -- & -- & N & N & N & N & 2 \\
J032901.63+312018.6 & -- & Class I & N & N & N & N & 1 \\
J032901.96+311538.0 & -- & -- & N & F & N & N & 2 \\
J032903.38+311601.6 & -- & -- & Y & N & N & N & 2 \\
J032903.75+311603.7 & -- & Class II & Y & P & N & N & 5,6 \\
J032904.06+311446.2 & -- & Class I & N & N & N & N & 2,5 \\
J032904.26+311609.0 & -- & -- & Y & N & N & N & 2 \\
J032907.75+312157.1 & -- & Class I & N & P & Y & N & 1 \\
J032909.14+312144.0 & -- & Class III & Y & P & Y & N & 3 \\
J032910.22+312335.1 & -- & Class II & N & F & N & N & 5 \\
J032910.39+312159.0 & F5 & Class II & N & N & Y & Y & 5,9 \\
J032910.42+311332.0 & -- & Class 0 & N & P & N & N & 1 \\
J032910.53+311330.9 & -- & Class 0 & N & F & N & N & 1 \\
J032911.25+311831.1 & -- & Class 0 & N & P & N & N & 1 \\
J032916.59+311648.7 & -- & -- & Y & P & N & N & 2 \\
J032917.67+312244.9 & K2/3IIIe & Class II & N & P & Y & N & 3,6 \\
J032922.29+311354.2 & -- & Class I & N & N & N & Y & 5 \\
\hline
\multicolumn{8}{c}{IC 348} \\
\hline
J034351.23+321309.1 & G5 & Class III &  Y & N & Y & N & 7,10 \\
J034357.60+320137.3 & M0.5 & Class III &  Y & N & Y & Y & 7, 10 \\
J034359.65+320153.9 & M0.5 & Class II & Y & P & Y & N & 5,7 \\
J034416.78+320956.4 & K0 & -- & N & N & Y & Y & 7 \\
J034420.37+320158.4 & -- & Class I &  N & F & Y & N & 5 \\
J034421.56+321017.4 & M1.5 &  Class III & Y & P & Y & Y & 7,10 \\
J034421.67+320624.8 & M2.75 & Class III & Y & -- & Y & N & 7,10 \\
J034424.57+320357.5 & M1 & Class III & Y & N & Y & N & 7, 10 \\
J034427.03+320443.5 & M1 & Class III & Y & -- & Y & N & 7, 10 \\
J034432.60+320842.4 & M2.5 & Class III & Y & F & Y & Y & 7, 10 \\
J034432.77+320837.6 & G6 & Class III & N & N & Y & N & 7, 10 \\
J034434.87+320633.5 & K5.5 & Class III & N & N & Y & N & 7, 10 \\
J034436.93+320645.4 & G3 & Class II & Y & N & Y & Y & 5,7 \\
J034438.72+320841.9 & K3 & Class III & Y & N & Y & N & 7, 10 \\
J034439.17+320918.4 & G8 & Class III & Y & F & Y & N & 7, 10 \\
J034443.98+320135.2 & -- & Class 0 & N & F & N & N & 1 \\
J034450.64+321906.3 & A0 & Class III & Y & F & Y & N & 7, 10 \\
J034507.97+320401.6 & G4 & Class III & Y & F & Y & Y & 7, 10 

\enddata
\tablenotetext{a}{All Class III objects corresponding to the IC~348 region are actually classified as ``anemic" or ``star" by \citet{2006AJ....131.1574L}. This means that the contribution of infrared emission from the disk is very low or not existent.  Based on this, we argue this classification is equivalent to a Class III object.} 
\tablenotetext{b}{Var.\ = Y when the source variability is higher than 50\% in at least one frequency; N when it is lower. $\alpha$ refers to the spectral index, and is given as P (for positive) when it is higher than 0.2; F (for flat) when it is between --0.2 and $+$0.2, and N (for negative) when is is lower than --0.2. X-ray\ = Y when there is a X-ray flux reported in literature, N when it is not.}
\tablenotetext{c}{VLBI Candidates = Sources that might have non-thermal emission and have high enough flux density to permit VLBI parallax measurements, Y when source is candidate, N when it is not.}
\tablenotetext{d}{1 = \citet{2009ApJ...692..973E}; 2 = \citet{1997ApJ...480L.125R}. \citet{1999ApJS..125..427R}. \citet{2000ApJ...542L.123A}; 3 = \citet{2010AJ....140..266W}; 4 = \citet{2009ApJS..181..321E}; 5 = \citet{2009ApJS..184...18G}. \citet{2008ApJ...674..336G}; 6 = \citet{1980AJ.....85.1638T}; 7 = \citet{2011ApJ...727...64K}; 8 = \citet{2011ApJ...736...25F}; 9 = \citet{2010AJ....140.1214C}. \citet{2008AJ....135.2496C}; 10 = \citet{2006AJ....131.1574L}}
\end{deluxetable}

\clearpage

\begin{deluxetable}{cccc}
\tabletypesize{\scriptsize}
\tablewidth{0pt}
\tablecolumns{4}
\tablecaption{YSOs candidates based only on their radio properties \label{tab:candidates}}
\tablehead{GBS--VLA Name & Variability$_{4.5\textrm{GHz}}$ & Variability$_{7.5\textrm{GHz}}$ & Spectral Index \\
{} & \colhead{(\%)} & \colhead{(\%)} & {} \\}
\startdata 
\multicolumn{4}{c}{NGC 1333} \\
\hline
J032813.80+311755.1 & $>$69 $\pm$ 11 & -- & -- \\
J032825.98+311616.0 & 52 $\pm$ 11 & $>$18 $\pm$ 35 & -3.0 $\pm$ 0.5 \\
J032846.49+312943.5 & $>$87 $\pm$ 4 & -- & -- \\
J032907.87+312348.0 & $>$57 $\pm$ 13 & $>$37 $\pm$ 29 & -1.4 $\pm$ 1.0 \\
J032918.56+311427.3 & 76 $\pm$ 15 & -- & -- \\
J032933.19+312845.2 & $>$71 $\pm$ 24 & -- & $<$-0.2 $\pm$ 0.5 \\
J032944.99+312019.7 & $>$58 $\pm$ 18 & -- & $<$-1.8 $\pm$ 0.4 \\ 
\hline
\multicolumn{4}{c}{IC 348} \\
\hline
J034347.85+320555.2 & $>$63 $\pm$  21 & -- & -- \\
J034358.35+315754.7 & $>$66 $\pm$ 21 & -- & -- \\
J034434.05+320104.3 & $>$82 $\pm$  14 & -- & -- \\
J034437.73+321839.3 & $>$76 $\pm$   4 & -- & -- \\
J034439.42+320128.8 & $>$71 $\pm$  17 & -- & -- \\
J034446.97+321455.6 & 52 $\pm$   7 & 34 $\pm$  11 & -0.9 $\pm$  0.2 \\
J034447.02+321457.9 & 43 $\pm$  10 & $>$61 $\pm$   9 & -1.3 $\pm$  0.4 \\ 
\hline
\multicolumn{4}{c}{Single Fields} \\
\hline
J032827.62+304909.4 & $>$94 $\pm$   2 & -- & -- \\
J032841.15+304945.2 & $>$92 $\pm$   2 & -- & -- \\
J032912.84+304558.5 & $>$65 $\pm$  10 & -- & -- \\
J032919.25+304548.7 & $>$96 $\pm$   2 & -- & -- \\
J033100.76+313412.3 & 59 $\pm$  15 & -- & -- \\
J033517.65+311650.0 & 50 $\pm$  15 & $>$28 $\pm$  38 &  1.0 $\pm$  1.2 \\ 
\enddata
\end{deluxetable}

\clearpage

\begin{deluxetable}{lccccc}
\tabletypesize{\scriptsize}
\tablewidth{0pt}
\tablecolumns{6}
\tablecaption{Proper motions of YSOs in Perseus \label{tab:motions}}
\tablehead{\colhead{GBS-VLA Name} & \colhead{ID (lit.)} & \colhead{$\mu_\alpha \cos(\delta)$} & \colhead{$\mu_\delta$}  & \colhead{$\mu$} & \colhead{$v_T$}  \\
{} & {} &\colhead{(mas yr$^{-1}$)} & \colhead{(mas yr$^{-1}$)} & \colhead{(mas yr$^{-1}$)} & \colhead{(km s$^{-1}$)} }
\startdata
GBS-VLA J032901.96+311538.1 & VLA 2 & 6.72 $\pm$ 0.69 & -13.96 $\pm$ 1.8 & 15 $\pm$ 2 & 16 $\pm$ 3 \\
GBS-VLA J032903.38+311601.6 & VLA 3 & 7.34 $\pm$ 3.26 & -15.87 $\pm$ 1.14 & 17 $\pm$ 3 & 19 $\pm$ 3 \\
Weighted Mean & {} & 6.74 $\pm$ 0.67 & -15.32 $\pm$ 0.92 & 17 $\pm$ 2 & 19 $\pm$ 2 

\enddata
\end{deluxetable}

\end{document}